\documentclass[12pt,a4paper]{article}
\pdfoutput=1

\usepackage[a4paper]{geometry}
\usepackage{amsmath}
\usepackage{amsfonts}
\usepackage{amssymb}
\usepackage{dsfont}
\usepackage[nosort]{cite}
\usepackage[backref=page]{hyperref}
\usepackage{graphicx}
\usepackage[margin=20pt,font=small,labelfont=bf]{caption}
\usepackage[margin=0pt,font=small,labelfont=normalfont]{subcaption}
\usepackage{enumitem}
\usepackage{multirow}
\usepackage{diagbox}

\newcommand{\eq}[1]{\begin{equation}
                     \begin{split} #1 \end{split}
                     \end{equation}}
\newcommand{\ov}{\overline}
\newcommand{\op}{\hspace{1pt}}

\allowdisplaybreaks[2]
\numberwithin{equation}{section}    

\begin{document}

\vspace*{\stretch{1}}

\begin{center}
{\LARGE
Flux vacua of the mirror octic  \\
}
\end{center}

\vspace{0.6cm}

\begin{center}
Erik Plauschinn, Lorenz Schlechter
\end{center}

\vspace{0.6cm}

\begin{center} 
\textit{
Institute for Theoretical Physics, Utrecht University \\
Princetonplein 5, 3584CC Utrecht \\
The Netherlands \\
}
\end{center} 

\vspace{2cm}

%%%%%%%%%%%%%%%%%%%%%%%%%%%%%%%%%%%%%%%%%%%%%%%
%%%%%%%%%%%%%%%%%%%%%%%%%%%%%%%%%%%%%%%%%%%%%%%
%%%%%%%%%%%%%%%%%%%%%%%%%%%%%%%%%%%%%%%%%%%%%%%
%%%%%%%%%%%%%%%%%%%%%%%%%%%%%%%%%%%%%%%%%%%%%%%

\begin{abstract}
\noindent
We determine \textit{all} flux vacua 
with flux numbers $N_{\rm flux}\leq 10$
for a  type IIB orientifold-compactification 
on the mirror-octic three-fold.
To achieve this, we develop and apply techniques for performing a complete scan of flux vacua for
the whole moduli space
--- 
we do not randomly sample fluxes nor do we consider only 
boundary regions of the moduli space.
We compare our findings to results in the literature.
\end{abstract}

\vspace*{\stretch{1}}

%%%%%%%%%%%%%%%%%%%%%%%%%%%%%%%%%%%%%%%%%%%%%%%
%%%%%%%%%%%%%%%%%%%%%%%%%%%%%%%%%%%%%%%%%%%%%%%
%%%%%%%%%%%%%%%%%%%%%%%%%%%%%%%%%%%%%%%%%%%%%%%
%%%%%%%%%%%%%%%%%%%%%%%%%%%%%%%%%%%%%%%%%%%%%%%
%%%%%%%%%%%%%%%%%%%%%%%%%%%%%%%%%%%%%%%%%%%%%%%
%%%%%%%%%%%%%%%%%%%%%%%%%%%%%%%%%%%%%%%%%%%%%%%
%%%%%%%%%%%%%%%%%%%%%%%%%%%%%%%%%%%%%%%%%%%%%%%
%%%%%%%%%%%%%%%%%%%%%%%%%%%%%%%%%%%%%%%%%%%%%%%

\clearpage
\tableofcontents

%%%%%%%%%%%%%%%%%%%%%%%%%%%%%%%%%%%%%%%%%%%%%%%
%%%%%%%%%%%%%%%%%%%%%%%%%%%%%%%%%%%%%%%%%%%%%%%
%%%%%%%%%%%%%%%%%%%%%%%%%%%%%%%%%%%%%%%%%%%%%%%
%%%%%%%%%%%%%%%%%%%%%%%%%%%%%%%%%%%%%%%%%%%%%%%
%%%%%%%%%%%%%%%%%%%%%%%%%%%%%%%%%%%%%%%%%%%%%%%
%%%%%%%%%%%%%%%%%%%%%%%%%%%%%%%%%%%%%%%%%%%%%%%
%%%%%%%%%%%%%%%%%%%%%%%%%%%%%%%%%%%%%%%%%%%%%%%
%%%%%%%%%%%%%%%%%%%%%%%%%%%%%%%%%%%%%%%%%%%%%%%
%%%%%%%%%%%%%%%%%%%%%%%%%%%%%%%%%%%%%%%%%%%%%%%
%%%%%%%%%%%%%%%%%%%%%%%%%%%%%%%%%%%%%%%%%%%%%%%
%%%%%%%%%%%%%%%%%%%%%%%%%%%%%%%%%%%%%%%%%%%%%%%
%%%%%%%%%%%%%%%%%%%%%%%%%%%%%%%%%%%%%%%%%%%%%%%

\clearpage
\section{Introduction}

String theory is a theory of quantum gravity  defined in ten space-time dimensions. 
To make contact with four-dimensional physics 
one usually compactifies the theory 
on a six-dimensional manifold, where typical compactification spaces are orientifolds of
Calabi-Yau three-folds. 
The  geometry of the compact space then determines properties of the four-dimensional 
theory, in particular, the topology determines the number of
massless scalar fields (moduli) in four dimensions.
Such fields are undesirable for phenomenological reasons. 
However, by considering non-vanishing vacuum-expectation-values for $p$-form gauge fields (fluxes)
one can deform the compact space and generate mass-terms 
for the moduli --- this  is called moduli stabilization.

A suitable framework for studying moduli stabilization in string theory is that of type IIB 
orientifold compactifications 
with O3- and O7-planes. 
The axio-dilaton and complex-structure moduli of the closed-string sector
can be stabilized by 
Neveu-Schwarz--Neveu-Schwarz (NS-NS) and 
Ramond-Ramond (R-R) three-form 
fluxes  \cite{Dasgupta:1999ss,Giddings:2001yu}, while the  K\"ahler moduli 
can be stabilized by non-perturbative effects via the 
KKLT \cite{Kachru:2003aw} or Large-Volume Scenario~\cite{Balasubramanian:2005zx}.

%%%%%%%%%%%%%%%%%%%%%%%%%%%%%%%%%%%%%%%%%%%%%%%
%%%%%%%%%%%%%%%%%%%%%%%%%%%%%%%%%%%%%%%%%%%%%%%

\subsubsection*{Finiteness of the landscape}

The effective theories obtained from compactifying string theory are said to belong to the 
string-theory landscape \cite{Susskind:2003kw}. 
Naively one might expect that there are infinitely-many compactification spaces 
as well as infinitely-many flux configurations and, hence,  
 there should be  infinitely-many effective theories 
originating from  string theory. 
However, there is evidence that  the string-theory landscape is finite
\cite{
Vafa:2005ui,
Acharya:2006zw,
Grimm:2021vpn
}. More concretely, 
\begin{itemize}

\item 
in regard to the number of topologically distinct Calabi-Yau three-folds 
we note that almost all known constructions 
are bi-rationally equivalent to descriptions that admit an elliptic fibration 
and that the number of topological types of those is finite  
\cite{Grassi1991,Gross:1993fd}.
We refer for instance to section~1 of \cite{Jejjala:2022lxh} for a
 brief overview of this matter.

\item The finiteness of the number of flux vacua has been investigated 
by deriving and analyzing a flux density
in the series of papers
\cite{
Ashok:2003gk,
Denef:2004ze,
Douglas:2006zj,
Lu:2009aw
}.
These results have been checked and verified in a number of concrete examples, for 
instance in 
\cite{
DeWolfe:2004ns,
Giryavets:2004zr,
Conlon:2004ds,
Eguchi:2005eh,
Braun:2011av
}.
More recently it has been proven in 
\cite{
Grimm:2020cda,
Bakker:2021uqw
}
that for an upper bound on the  flux number $N_{\rm flux}$ (to be defined below)
the number of self-dual flux configurations in F-theory is finite.
This  extends to imaginary self-dual fluxes in the type IIB orientifold limit mentioned above.

\end{itemize}

%%%%%%%%%%%%%%%%%%%%%%%%%%%%%%%%%%%%%%%%%%%%%%%
%%%%%%%%%%%%%%%%%%%%%%%%%%%%%%%%%%%%%%%%%%%%%%%

\subsubsection*{Motivation}

Our motivation for the present work is to explicitly show the finiteness of flux vacua for a concrete example.
The finiteness proof in \cite{Bakker:2021uqw} is not constructive and does not 
provide a procedure to determine all flux vacua for a given flux number. 
In the context of type IIB orientifold compactifications with  NS-NS and R-R three-form fluxes,
we therefore ask the question
\begin{quote}
\textit{Can we construct all  flux vacua for a given flux number in a simple example?}
\end{quote}
Our emphasis here is on \textit{all} flux vacua --- that is, we do not want to randomly sample flux configurations 
nor do we want to consider only a particular region in moduli space.

The results obtained from such a complete scan of  vacua can be valuable in 
different ways. For instance,
1) one can test explicitly the estimated behavior  of  the number of flux 
vacua made in \cite{Ashok:2003gk,Denef:2004ze}. 
2) Since the flux landscape is often explored by randomly sampling flux configurations
(see e.g.~\cite{Dubey:2023dvu} for recent work on this matter), 
it is possible to analyze to what degree such  sampling  approaches can capture the majority of vacua.
3) One can furthermore use such data to investigate swampland conjectures like the 
tadpole conjecture \cite{Bena:2020xrh} or explore the desert region 
of the moduli space~\cite{Long:2021jlv}.

%%%%%%%%%%%%%%%%%%%%%%%%%%%%%%%%%%%%%%%%%%%%%%%
%%%%%%%%%%%%%%%%%%%%%%%%%%%%%%%%%%%%%%%%%%%%%%%

\subsubsection*{The mirror octic}

To address the above question, we develop and apply techniques to determine all flux vacua for 
a type IIB orientifold projection on the mirror-octic three-fold.
This model has one complex-structure modulus and  has been studied before via statistical methods in 
\cite{DeWolfe:2004ns,Giryavets:2004zr} 
and more recently using machine-learning techniques in \cite{Cole:2018emh,Cole:2019enn}.
However, in these works only certain regions of moduli space have been considered 
and no complete scan for vacua has been performed.

We also note that for the mirror octic  there exists an orientifold projection together with a configuration of D-branes 
such that the D3-brane tadpole contribution coming from orientifold planes and D7-branes is 
$Q_{\rm D3}=8$ (c.f.~\cite{Moritz:2023jdb}). 
Via the tadpole-cancellation condition this implies that the flux number can at most be eight.
For the scan we perform in this paper we consider flux numbers $N_{\rm flux}\leq 10$, and 
hence we classify all flux configurations for this model.

%%%%%%%%%%%%%%%%%%%%%%%%%%%%%%%%%%%%%%%%%%%%%%%
%%%%%%%%%%%%%%%%%%%%%%%%%%%%%%%%%%%%%%%%%%%%%%%

\subsubsection*{Outline}

This paper is organized as follows: 
in section~\ref{sec_comp_iib} we introduce our setting and notation.
The reader familiar with moduli stabilization in type IIB orientifolds can skip this part. 
In section~\ref{sec_flux_vacua} we explain our general strategy for determining a
finite set of flux configurations; 
we emphasize that this discussion is valid for an arbitrary number 
of complex-structure moduli. 
In section~\ref{sec_octic} we outline  some technical details for computing the 
period vector of the mirror octic
and in section~\ref{sec_strategy} we describe the details of our scan for vacua. 
These two sections contain the technical details of our work and can be skipped by 
the reader not interested in those.
In section~\ref{sec_results} we present and discuss our results 
and section~\ref{sec_conclusion} contains our conclusions.

%%%%%%%%%%%%%%%%%%%%%%%%%%%%%%%%%%%%%%%%%%%%%%%
%%%%%%%%%%%%%%%%%%%%%%%%%%%%%%%%%%%%%%%%%%%%%%%
%%%%%%%%%%%%%%%%%%%%%%%%%%%%%%%%%%%%%%%%%%%%%%%
%%%%%%%%%%%%%%%%%%%%%%%%%%%%%%%%%%%%%%%%%%%%%%%
%%%%%%%%%%%%%%%%%%%%%%%%%%%%%%%%%%%%%%%%%%%%%%%
%%%%%%%%%%%%%%%%%%%%%%%%%%%%%%%%%%%%%%%%%%%%%%%
%%%%%%%%%%%%%%%%%%%%%%%%%%%%%%%%%%%%%%%%%%%%%%%
%%%%%%%%%%%%%%%%%%%%%%%%%%%%%%%%%%%%%%%%%%%%%%%
%%%%%%%%%%%%%%%%%%%%%%%%%%%%%%%%%%%%%%%%%%%%%%%
%%%%%%%%%%%%%%%%%%%%%%%%%%%%%%%%%%%%%%%%%%%%%%%

\clearpage
\section{Type IIB flux compactifications}
\label{sec_comp_iib}

We start by briefly reviewing moduli stabilization 
for type IIB orientifold compactifications
with Neveu-Schwarz--Neveu-Schwarz and Ramond-Ra\-mond three-form fluxes.
The purpose of this section is to  establish our notation and conventions ---
the reader familiar with this topic can safely skip to the next section.

%%%%%%%%%%%%%%%%%%%%%%%%%%%%%%%%%%%%%%%%%%%%%%%
%%%%%%%%%%%%%%%%%%%%%%%%%%%%%%%%%%%%%%%%%%%%%%%

\subsubsection*{Moduli}

We consider orientifold compactifications of type IIB string theory on Calabi-Yau three-folds $\mathcal X$, 
where the orientifold projection is chosen such that its fixed loci are
O3- and O7-planes. Due to this projection,  the cohomologies 
of   $\mathcal X$ are split into even and odd eigenspaces 
$H^{p,q}_{\pm}(\mathcal X)$ whose dimensions will be denoted by $h^{p,q}_{\pm}$.
The effective four-dimensional theory obtained after compactification 
contains massless scalar fields, in particular,
the axio-dilaton $\tau$, $h^{2,1}_-$ complex-structure moduli $z^i$, 
 $h^{1,1}_+$ K\"ahler moduli $T_a$, and  $h^{1,1}_-$ moduli $G_{\hat a}$.
We parametrize the axio-dilaton as
\eq{
\label{coord_tau}
\tau = c+ i \op s \,, 
}
and the physical region of the dilaton is characterized by $s>0$.
The K\"ahler potential for these  fields  at leading order in $\alpha'$ and the string coupling is given by 
\eq{
\label{kpot}
  \mathcal K =  - \log\bigl[ -i(\tau-\bar \tau) \bigr] - \log \left[ +i\int_{\mathcal X} \Omega \wedge \bar \Omega \right]
  - 2\log \mathcal V\,,
}
where $\Omega$ denotes the holomorphic three-form of $\mathcal X$, 
which depends on the complex-structure moduli $z^i$,
and $\mathcal V$ denotes the volume of $\mathcal X$,  which 
depends on the K\"ahler moduli $T_a$ and on the moduli $G_{\hat a}$. 

%%%%%%%%%%%%%%%%%%%%%%%%%%%%%%%%%%%%%%%%%%%%%%%
%%%%%%%%%%%%%%%%%%%%%%%%%%%%%%%%%%%%%%%%%%%%%%%

\subsubsection*{Fluxes}

In order to stabilize the axio-dilaton and complex-structure moduli we 
consider non-vanishing 
NS-NS and R-R
three-form fluxes 
$H$ and $F$ (for reviews see \cite{Grana:2005jc,Douglas:2006es}). 
The fluxes are integer quantized and can be expanded 
in an integral 
symplectic basis $\{\alpha_I,\beta^I\}\in H^3_-(\mathcal X,\mathbb Z)$ as
\eq{
  \label{fluxes_200}
  H = h^I \alpha_I + h_I \op\beta^I\,,
  \hspace{50pt}
  F = f^I \alpha_I + f_I \op\beta^I\,,  
}
where $h^I,h_I,f^I,f_I\in\mathbb Z$ and 
$I=0,\ldots,h^{2,1}_-$.
These fluxes generate a scalar potential for the four-dimensional theory that 
can be computed from  the superpotential
\eq{
\label{wpot}
  W = \int_{\mathcal X} \Omega \wedge G \,,
  \hspace{50pt}
  G = F -\tau \op H \,.
}

%%%%%%%%%%%%%%%%%%%%%%%%%%%%%%%%%%%%%%%%%%%%%%%
%%%%%%%%%%%%%%%%%%%%%%%%%%%%%%%%%%%%%%%%%%%%%%%

\subsubsection*{Tadpole cancellation condition}

The fluxes appearing in the superpotential \eqref{wpot} are constrained by the tadpole cancellation conditions. 
One of these conditions takes the schematic form
\eq{
\label{tadpole_d3}
0=N_{\rm flux} +  2\op N_{{\rm D}3}  + Q_{\rm D3} \,,
}
where $N_{{\rm D}3}$ denotes the number of space-time filling D3-branes, $Q_{\rm D3}$ 
denotes the contribution from O3- and O7-planes and from D7-branes, and 
$N_{\rm flux}$ denotes the flux number defined as
\eq{
  \label{tad_03}
  N_{\rm flux} &= \int_{\mathcal X} F \wedge H  \,.
}
For details on the condition \eqref{tadpole_d3}
see for instance equation (3.6c) in \cite{Plauschinn:2020ram}.
Note that 
$N_{\rm flux}$ is bounded from below and from above
because 1)
$Q_{\rm D3}$ is bounded from below and typically negative,
and 2) $N_{\rm flux}$ has to be positive in order for the dilaton $s$ to be positive.

%%%%%%%%%%%%%%%%%%%%%%%%%%%%%%%%%%%%%%%%%%%%%%%
%%%%%%%%%%%%%%%%%%%%%%%%%%%%%%%%%%%%%%%%%%%%%%%

\subsubsection*{F-term minima of the scalar potential}

Since the K\"ahler potential \eqref{kpot} that we  consider is of no-scale type and
because the superpotential \eqref{wpot} does not depend
on the  moduli $T_a$ or  $G_{\hat a}$, the standard F-term scalar potential 
can be brought into the form
\eq{
\label{spot}
  V = e^{\mathcal K }  F_{\vphantom{\ov \beta}\alpha} \op \mathcal G^{\alpha\ov \beta}\op \ov F_{\ov \beta} \,.
}
The K\"ahler potential $\mathcal K$ was shown in \eqref{kpot},
the F-terms are given by the K\"ahler-covariant derivative of $W$ as $F_{\alpha} = \partial_{\alpha} W + (\partial_{\alpha} \mathcal K) W$ with $\alpha=(\tau,z^i)$,
and $\mathcal G^{\alpha\ov \beta}$ denotes the inverse of the K\"ahler metric
$\mathcal G_{\alpha \ov \beta} = \partial_{\vphantom{\ov \beta}\alpha} \partial_{\ov \beta}\op \mathcal K$.
We are interested in global minima of the scalar potential \eqref{spot} which are given by
the vanishing F-terms
\eq{
\label{min_99}
 F_{\tau}=0\,, \hspace{50pt}F_{z^i}=0\,.
}
Using the Hodge-star operator $\star$ on the compact manifold $\mathcal X$, 
these F-term conditions  are equivalent to the 
following imaginary self-duality condition on the three-form flux $G$ \cite{Giddings:2001yu}
\eq{
  \label{min_100}
  G =- i\star  G\, .
}

%%%%%%%%%%%%%%%%%%%%%%%%%%%%%%%%%%%%%%%%%%%%%%%
%%%%%%%%%%%%%%%%%%%%%%%%%%%%%%%%%%%%%%%%%%%%%%%
%%%%%%%%%%%%%%%%%%%%%%%%%%%%%%%%%%%%%%%%%%%%%%%
%%%%%%%%%%%%%%%%%%%%%%%%%%%%%%%%%%%%%%%%%%%%%%%
%%%%%%%%%%%%%%%%%%%%%%%%%%%%%%%%%%%%%%%%%%%%%%%
%%%%%%%%%%%%%%%%%%%%%%%%%%%%%%%%%%%%%%%%%%%%%%%
%%%%%%%%%%%%%%%%%%%%%%%%%%%%%%%%%%%%%%%%%%%%%%%
%%%%%%%%%%%%%%%%%%%%%%%%%%%%%%%%%%%%%%%%%%%%%%%
%%%%%%%%%%%%%%%%%%%%%%%%%%%%%%%%%%%%%%%%%%%%%%%
%%%%%%%%%%%%%%%%%%%%%%%%%%%%%%%%%%%%%%%%%%%%%%%

\clearpage
\section{Flux vacua and where to find them}
\label{sec_flux_vacua}

Although the flux number $N_{\rm flux}$ is bounded by the tadpole cancellation condition \eqref{tadpole_d3}, 
a priori there are  infinitely-many  choices for  $H$ and $F$ that satisfy such a bound.
However, as  has been argued for and proven in 
\cite{
Ashok:2003gk,
Denef:2004ze
}
and
\cite{
Grimm:2020cda,
Bakker:2021uqw
},
for a given $N_{\rm flux}$ the number of physically-inequivalent type IIB flux vacua  is finite. 
Our objective for this section is to make this result explicit for  a finite region in  complex-structure moduli space, where finite means that the region does not contain any singularities such as the conifold or the large-complex-structure 
point.
In particular, we  derive  bounds on the fluxes $H$ and $F$
that can be implemented in computer searches.

%%%%%%%%%%%%%%%%%%%%%%%%%%%%%%%%%%%%%%%%%%%%%%%
%%%%%%%%%%%%%%%%%%%%%%%%%%%%%%%%%%%%%%%%%%%%%%%
%%%%%%%%%%%%%%%%%%%%%%%%%%%%%%%%%%%%%%%%%%%%%%%
%%%%%%%%%%%%%%%%%%%%%%%%%%%%%%%%%%%%%%%%%%%%%%%
%%%%%%%%%%%%%%%%%%%%%%%%%%%%%%%%%%%%%%%%%%%%%%%

\subsection{Matrix notation}
\label{sec_matrix}

For our  discussion below it will be convenient to work in matrix notation. 
In this subsection we therefore translate the objects appearing in section~\ref{sec_comp_iib}
into vectors and matrices.

%%%%%%%%%%%%%%%%%%%%%%%%%%%%%%%%%%%%%%%%%%%%%%%
%%%%%%%%%%%%%%%%%%%%%%%%%%%%%%%%%%%%%%%%%%%%%%%

\subsubsection*{Flux and period vectors}

We first recall the expansions of the three-form fluxes $H$ and $F$ shown in equation \eqref{fluxes_200}.
The integer coefficients in these expansions can be combined into 
two \mbox{$2(h^{2,1}_-+1)$}-dimensional integer flux vectors as follows
\eq{
\label{rel_270}
  \mathsf H = \binom{h^I}{h_I}\,,
  \hspace{60pt}
  \mathsf F = \binom{f^I}{f_I}\,.
}
For  the holomorphic three-form $\Omega$ and its K\"ahler-covariant derivative
$D_{z^i}\Omega = \partial_{z^i} \Omega+ (\partial_{z^i} \mathcal K)\,\Omega$
we perform an expansion
similar to
 \eqref{fluxes_200}
 using the integral symplectic basis 
$\{\alpha_I,\beta^I\}\in H^3_-(\mathcal X,\mathbb Z)$. We write
\eq{
\label{omega}
  \Omega= \Pi^J \alpha_J+ \Pi_J \beta^J \,,
  \hspace{60pt}
  D_{z^i}\Omega= D_i\Pi^J \alpha_J+ D_i\Pi_J \beta^J \,,
}
and  define the period vector $\Pi$ and its covariant derivative $D_i\Pi$ as
\eq{
  \Pi = \binom{\Pi^J}{\Pi_J}\,,
  \hspace{60pt}
  D_i\Pi = \binom{D_i\Pi^J}{D_i\Pi_J}
   \,.
}

%%%%%%%%%%%%%%%%%%%%%%%%%%%%%%%%%%%%%%%%%%%%%%%
%%%%%%%%%%%%%%%%%%%%%%%%%%%%%%%%%%%%%%%%%%%%%%%

\subsubsection*{Symplectic pairing and Hodge-star matrix}

The symplectic pairing for the basis $\{\alpha_I,\beta^I\}$ can  be represented by the following
$2(h^{2,1}_-+1)\times 2(h^{2,1}_-+1)$-dimensional anti-symmetric 
matrix
\eq{
   \eta = \int_{\mathcal X} \binom{\alpha}{\beta} \wedge  \bigl(\alpha,\beta\bigr) 
  =
   \arraycolsep2pt
   \left( \begin{array}{cc}
   0    & + \mathds 1
  \\[2pt]
   -\mathds 1 & 0
  \end{array}\right) .
}
The matrix representing the Hodge-star operator $\star$
can be expressed in terms of the period vector 
$\Pi$ and the conjugate of $D_{i}\Pi$. 
In particular, we introduce the following two $(h^{2,1}_-+1)\times (h^{2,1}_-+1)$-dimensional matrices
\eq{
  P{}_{IJ} = \bigl( \Pi_I,\op \ov{D_1\Pi}_I, \op \ov{D_2\Pi}_I, \op\ldots \bigr)\,,
  \hspace{40pt}
  Q{}^I{}_{J} = \bigl( \Pi^I,\op \ov{D_1\Pi}^I, \op \ov{D_2\Pi}^I, \op\ldots \bigr)\,,
}
which we use to define the period matrix
$\mathcal N_{IJ}$  as 
\eq{
  \label{gkf}
  \mathcal N_{IJ} 
  = \mathcal R_{IJ} + i\op \mathcal I_{IJ}
  = -P_{IM}(Q^{-1}){}^M{}_J\,.
}
Note that this expression agrees with the maybe more-familiar formula in case a pre-potential exists
(see e.g.~equation (2.15) in \cite{Tsagkaris:2022apo}).
Using the period matrix $\mathcal N$, the Hodge-star operator can be represented by the 
following 
$2(h^{2,1}+1)\times 2(h^{2,1}+1)$-dimensional matrix
\eq{
\label{mat_197}
   \mathcal M = \int_{\mathcal X} \binom{\alpha}{\beta} \wedge \star \bigl(\alpha,\beta\bigr) 
  =
   \arraycolsep3pt
   \left( \begin{array}{cc}
   -\mathcal I - \mathcal R\op \mathcal I^{-1} \mathcal R
   &
   -\mathcal R\op \mathcal I^{-1} 
  \\[2pt]
   -\mathcal I^{-1}\op \mathcal R & - \mathcal I^{-1} 
  \end{array}\right) .
}
Note that this matrix is symmetric ($\mathcal M^T=\mathcal M$) and symplectic ($\mathcal M^T \eta\op\mathcal M = \eta$)
and it is required to be positive-definite. 
This implies that $\mathcal I$ needs to be negative-definite and, due to $\mathcal M$ being symplectic, that
the real eigenvalues of $\mathcal M$ come in pairs of the form
\eq{
\label{mat_198}
  \left(\lambda^{\vphantom{-1}}_I,\op \lambda^{-1}_I \right),
  \hspace{60pt}\lambda_I \geq 1\,.
}

%%%%%%%%%%%%%%%%%%%%%%%%%%%%%%%%%%%%%%%%%%%%%%%
%%%%%%%%%%%%%%%%%%%%%%%%%%%%%%%%%%%%%%%%%%%%%%%
%%%%%%%%%%%%%%%%%%%%%%%%%%%%%%%%%%%%%%%%%%%%%%%
%%%%%%%%%%%%%%%%%%%%%%%%%%%%%%%%%%%%%%%%%%%%%%%
%%%%%%%%%%%%%%%%%%%%%%%%%%%%%%%%%%%%%%%%%%%%%%%

\subsection{Minimum conditions}

We now turn to the conditions for the axio-dilaton and complex-structure moduli
that describe the global minimum of the scalar potential of the effective four-dimensional theory.

%%%%%%%%%%%%%%%%%%%%%%%%%%%%%%%%%%%%%%%%%%%%%%%
%%%%%%%%%%%%%%%%%%%%%%%%%%%%%%%%%%%%%%%%%%%%%%%

\subsubsection*{F-term conditions}

The global minimum of the scalar potential \eqref{spot} is given by vanishing F-terms. These  can be expressed in the following way
\eq{
  \arraycolsep2pt
  \begin{array}{lclclrcr}
    0 &=& \displaystyle  F_\tau &=&
    \displaystyle   \int_{\mathcal X} & \displaystyle \Omega \wedge \ov G  &=&
    \displaystyle  \Pi^T \eta \op     \bigl( \mathsf F-\ov\tau\op \mathsf H\bigr) \,,
  \\[12pt]
    0 &=& \displaystyle  F_{z^i}  &=&
    \displaystyle   \int_{\mathcal X} & \displaystyle D_{z^i}\Omega \wedge  G  &=& 
    \displaystyle D_i\Pi^T \eta \op 
    \bigl( \mathsf F-\tau\op \mathsf H\bigr) \,,
  \end{array}
}
and the first of these conditions can be solved for the axio-dilaton $\tau$ as follows
\eq{
\label{rel_402}
   \tau = \frac{\ov \Pi{}^T\eta \,\mathsf  F}{\ov\Pi{}^T\eta \, \mathsf H}\,.
}
Using this solution the remaining F-term conditions $F_{z^i}=0$ can be 
expressed using 
a real anti-symmetric  $2(h^{2,1}_-+1)\times 2(h^{2,1}_-+1)$-dimensional matrix $\rho$ as
\eq{
\label{rel_401}
  0 = D_i\Pi^T  \rho\, \ov\Pi\,,
  \hspace{50pt}
 \rho = \eta^T \frac{\mathsf H \op\mathsf F^T- \mathsf F \op \mathsf H^T}{N_{\rm flux}} \op\eta \,.
}
Note also that $\eta\op\rho$ is a projection matrix, that is $ (\eta\op \rho)^2 = \eta\op \rho$.

%%%%%%%%%%%%%%%%%%%%%%%%%%%%%%%%%%%%%%%%%%%%%%%
%%%%%%%%%%%%%%%%%%%%%%%%%%%%%%%%%%%%%%%%%%%%%%%

\subsubsection*{Superpotential at the minimum}

An important quantity when discussing moduli stabilization by fluxes is the value $W_0$ of the superpotential \eqref{wpot} at the minimum. 
Due to the freedom to rescale the holomorphic three-form $\Omega$ by any holomorphic function,
$W_0$ itself is not well-defined.
However, an invariant quantity is obtained by multiplying $W_0$ by 
$e^{\mathcal K_{\rm cs}/2}$, where $\mathcal K_{\rm cs}$ denotes the 
K\"ahler potential for the complex-structure moduli.
For determining the gravitino mass it is furthermore 
useful to  multiply by $e^{\mathcal K_{\tau}/2}$, where 
$\mathcal K_{\tau}$ denotes  the  K\"ahler potential for the axio-dilaton. 
Using the solution \eqref{rel_402} we then determine
\eq{
  \label{superpot_200}
  e^{\mathcal K_{\rm cs}}\lvert W_0\rvert^2 &=  \frac{N_{\rm flux}^2}{ +i\, \Pi^T\eta\,\ov\Pi}\,\left\lvert \frac{\Pi^T\rho\,\ov\Pi}{\Pi^T\eta\,\mathsf H} \right\rvert^2
  \biggr\rvert_{\rm min}\,,
  \\[6pt]
e^{\mathcal K_{\tau}+\mathcal K_{\rm cs}}\lvert W_0\rvert^2 &= 
  -N_{\rm flux} \, \frac{\Pi^T\rho\,\ov\Pi}{\Pi^T\eta\,\ov\Pi}
  \biggr\rvert_{\rm min}
  \,,
  }
where the matrix $\rho$ was defined in \eqref{rel_401} and where the right-hand sides are 
understood to be evaluated at the minimum in complex-structure moduli space.
Note also that 
the imaginary part of $\Pi^T\rho\,\ov\Pi$ is required to be positive for $s>0$ and that 
in our conventions the imaginary part of $\Pi^T\eta\,\ov\Pi$ is negative. 

%%%%%%%%%%%%%%%%%%%%%%%%%%%%%%%%%%%%%%%%%%%%%%%
%%%%%%%%%%%%%%%%%%%%%%%%%%%%%%%%%%%%%%%%%%%%%%%

\subsubsection*{Self-duality condition}

We also recall that  the requirement of vanishing F-terms shown in \eqref{min_99} corresponds  to the self-duality condition 
condition \eqref{min_100}. The latter can be split into a real and imaginary part, which 
are equivalent to each other. Focussing for definiteness on the real part, we have in form notation
\eq{
\label{min_101}
  F- H\op c = -\star H\op s\,.
}
In matrix notation, this can be written in the following two ways
\eq{
  \label{min_204}
  \mathsf F = \bigl( \mathds 1\op c + \eta \op \mathcal M \op s\bigr) \mathsf H\,,
  \hspace{40pt}
  \mathsf H = \frac{1}{s^2+c^2}\bigl( \mathds 1\op c - \eta \op \mathcal M \op s\bigr) \mathsf F\,.
}
The distinct eigenvalues of the matrix $( \mathds 1\op c + \eta \op \mathcal M \op s)$ are $c\pm i\op s$
and  the determinant is 
\raisebox{0pt}[0pt][0pt]{$\det ( \mathds 1\op c + \eta \op \mathcal M \op s) = (c^2+s^2)^{h^{2,1}_-+1}$}.
For a non-vanishing axio-dilaton the matrix above is thus invertible,
implying that if $\mathsf H$ or $\mathsf F$ are zero also
$\mathsf F$ or $\mathsf H$ have to vanish, respectively. In this case 
 no axio-dilaton or complex-structure moduli 
are stabilized.

%%%%%%%%%%%%%%%%%%%%%%%%%%%%%%%%%%%%%%%%%%%%%%%
%%%%%%%%%%%%%%%%%%%%%%%%%%%%%%%%%%%%%%%%%%%%%%%
%%%%%%%%%%%%%%%%%%%%%%%%%%%%%%%%%%%%%%%%%%%%%%%
%%%%%%%%%%%%%%%%%%%%%%%%%%%%%%%%%%%%%%%%%%%%%%%
%%%%%%%%%%%%%%%%%%%%%%%%%%%%%%%%%%%%%%%%%%%%%%%

\subsection{The axio-dilaton}

The axio-dilaton $\tau=c+i\op s$ plays an important role for our  discussion
in section~\ref{sec_bounds}. 
We therefore make  the following two observations.

%%%%%%%%%%%%%%%%%%%%%%%%%%%%%%%%%%%%%%%%%%%%%%%
%%%%%%%%%%%%%%%%%%%%%%%%%%%%%%%%%%%%%%%%%%%%%%%

\subsubsection*{Relations}

Applying  $H\wedge\star$ and $H\wedge$ to the minimum condition \eqref{min_101}
and integrating over the three-fold $\mathcal X$,
one finds for the axion and for the dilaton
\eq{
\label{min_102}
c = \frac{\int H\wedge \star F}{\int H\wedge\star H}\,,
\hspace{70pt}
s = \frac{N_{\rm flux}}{\int H\wedge\star H} \,.
}
Furthermore, applying $F\wedge \star$ to \eqref{min_101}, integrating over the three-fold $\mathcal X$, 
and using the expressions in \eqref{min_102}, one obtains the relation
\eq{
\label{min_110}
 \int_{\mathcal X} F\wedge\star F  &=  (s^2 +c^2) \int_{\mathcal X} H\wedge\star H\,.
}

%%%%%%%%%%%%%%%%%%%%%%%%%%%%%%%%%%%%%%%%%%%%%%%
%%%%%%%%%%%%%%%%%%%%%%%%%%%%%%%%%%%%%%%%%%%%%%%

\subsubsection*{Fundamental domain}

Let us also recall that the axio-dilaton $\tau$ as well as the fluxes $H$ and $F$ 
transform under the duality group $SL(2,\mathbb Z)$. Theories related by this transformation 
are physically equivalent and we can restrict $\tau$ to a fundamental 
domain typically chosen as
\eq{
\label{tau_fundamental_d}
  \mathcal F_{\tau}=\left\{ 
-\frac{1}{2}\leq c \leq0,\, c^2+s^2 \geq 1 
\; \cup \;
0<c< +\frac12, \,c^2+s^2 > 1 
\right\}.
}
This means in particular that the dilaton $s$ is bounded from below, and the smallest value of 
$s$ is reached at $c=-1/2$. Hence, we have the bound
\eq{
  \label{rel_202}
  \frac{\sqrt{3}}{2} \leq s \,.
}

%%%%%%%%%%%%%%%%%%%%%%%%%%%%%%%%%%%%%%%%%%%%%%%
%%%%%%%%%%%%%%%%%%%%%%%%%%%%%%%%%%%%%%%%%%%%%%%
%%%%%%%%%%%%%%%%%%%%%%%%%%%%%%%%%%%%%%%%%%%%%%%
%%%%%%%%%%%%%%%%%%%%%%%%%%%%%%%%%%%%%%%%%%%%%%%
%%%%%%%%%%%%%%%%%%%%%%%%%%%%%%%%%%%%%%%%%%%%%%%

\subsection{Bounds on the fluxes}
\label{sec_bounds}

In this subsection we  derive  bounds on the flux quanta appearing in 
the $H$- and $F$-fluxes. 
More concretely, we choose a finite region of complex-structure moduli space in which the 
eigenvalues of the Hodge-star matrix $\mathcal M$ are finite and derive
bounds on the Euclidean norms of $\mathsf H$ and $\mathsf F$.

%%%%%%%%%%%%%%%%%%%%%%%%%%%%%%%%%%%%%%%%%%%%%%%
%%%%%%%%%%%%%%%%%%%%%%%%%%%%%%%%%%%%%%%%%%%%%%%

\subsubsection*{Bounds I}

Let us start with the expression $\int H\wedge \star H$. Using the matrix notation introduced in section~\ref{sec_matrix} this
can be written as
\eq{
  \int_{\mathcal X} H\wedge \star H = \mathsf H^T\! \mathcal M\op  \mathsf H\,.
}
Note that due to $\mathcal M$ being positive definite and due to $\mathsf H$ being required to be non-vanishing, this expression is strictly positive for our purposes. 
Using standard bounds on matrix norms  we then have
\eq{
  \label{rel_102}
  \frac{1}{\lambda_{\rm max}} \lVert \mathsf H \rVert^2\leq\mathsf H^T \!\mathcal M\op \mathsf H
  \leq  \lambda_{\rm max} \lVert \mathsf H \rVert^2\,,
}
where $\lVert \mathsf H \rVert^2= \mathsf H^T  \mathsf H$ is the  Euclidean-norm squared of $\mathsf H$, $\lambda_{\rm max}$ denotes the largest eigenvalue of the Hodge-star matrix $\mathcal M$, and 
the smallest eigenvalue of $\mathcal M$ is given by $\lambda_{\rm min} = 1/\lambda_{\rm max}$ due to  \eqref{mat_198}.
Combining then \eqref{min_102} with \eqref{rel_202} and \eqref{rel_102} we obtain
\eq{
  \label{rel_121}
 \lVert \mathsf H \rVert^2\leq \frac{2N_{\rm flux}\lambda_{\rm max}}{\sqrt{3}}\,.
}
This means that when the eigenvalues of the Hodge-star matrix $\mathcal M$ and the flux number $N_{\rm flux}$ are finite,
the number of allowed $H$-fluxes is finite. 
In a similar spirit, when using only \eqref{rel_102} in  \eqref{min_102} and noting that $ \lVert \mathsf H \rVert^2\geq 1$, 
we can derive bounds for the dilaton as
\eq{
  \frac{\sqrt{3}}{2} \leq s \leq  N_{\rm flux} \op \lambda_{\rm max}\,.
}

%%%%%%%%%%%%%%%%%%%%%%%%%%%%%%%%%%%%%%%%%%%%%%%
%%%%%%%%%%%%%%%%%%%%%%%%%%%%%%%%%%%%%%%%%%%%%%%

\subsubsection*{Bounds II}

We now turn to the $F$-flux. 
Starting from the matrix expression of
\eqref{min_102} and using the corresponding form of \eqref{min_110} we obtain
\eq{
  \label{rel_210111}
\frac{N^2_{\rm flux}}{s^2}=(\mathsf H^T \!\mathcal M\op \mathsf H)^2
=\mathsf H^T \!\mathcal M\op \mathsf H\, \frac{\mathsf F^T \!\mathcal M\op \mathsf F}{c^2+s^2}
\hspace{20pt}\Rightarrow\hspace{20pt}
\mathsf F^T \!\mathcal M\op \mathsf F = \frac{N^2_{\rm flux}}{\mathsf H^T \!\mathcal M\op \mathsf H}\frac{s^2+c^2}{s^2}\,.
}
Employing  then bounds of the form \eqref{rel_102}, 
the requirement $ \lVert \mathsf H \rVert^2\geq 1$, the relation on $s$ shown 
in \eqref{rel_202}, and $c^2\leq 1/4$, we arrive at
\eq{
  \label{rel_210}
 \lVert \mathsf F \rVert^2 \leq \frac{4N^2_{\rm flux}\lambda^2_{\rm max}}{3}\,. 
}
This relation implies again that for finite eigenvalues of the Hodge-star matrix 
$\mathcal M$ and for a finite flux number $N_{\rm flux}$ the number of $F$-flux quanta is finite.

%%%%%%%%%%%%%%%%%%%%%%%%%%%%%%%%%%%%%%%%%%%%%%%
%%%%%%%%%%%%%%%%%%%%%%%%%%%%%%%%%%%%%%%%%%%%%%%

\subsubsection*{Bounds III}

However, we can strengthen the bound shown in equation \eqref{rel_210}.
To do so, let us use the solution for $s$ shown in \eqref{min_102} to rewrite the relation \eqref{min_110} as
\eq{
  \label{rel_204}
  \int_{\mathcal X} F\wedge\star F  &=  \frac{N_{\rm flux}^2}{\int H\wedge\star H} +c^2\int_{\mathcal X} H\wedge\star H \,.
}
We then define the two quantities
\eq{
  \mathsf x = \frac{ \int H\wedge\star H}{N_{\rm flux}}\,, \hspace{50pt}
  \mathsf y = \frac{ \int F\wedge\star F}{N_{\rm flux}}\,,
}
and rewrite \eqref{rel_121} and \eqref{rel_204} as
\eq{
  \label{rel_205}
  \frac{\lVert \mathsf H \rVert^2}{\lambda_{\rm max}\op N_{\rm flux}} \leq \mathsf x\leq \frac{2}{\sqrt{3}}\,, \hspace{50pt}
  \mathsf y = \frac{1}{\mathsf x} + c^2\op \mathsf x \,.
}
The function $\mathsf y(\mathsf x)$ has a global minimum at $\mathsf x_{\rm min}= 1/\lvert c\rvert$, and since
$\lvert c\rvert \leq 1/2$ the minimum is in the region $\mathsf x_{\rm min} \geq 2$
and thus outside the allowed range for $\mathsf x$. In the range shown as
the first relation in \eqref{rel_205} 
the function $\mathsf y(\mathsf x)$ is therefore monotonically decreasing, which implies
\eq{
  \label{rel_2051}
  \mathsf y\left(\mathsf x=\frac{2}{\sqrt{3}}\right) \leq &\: \mathsf y \leq 
    \mathsf y\left(\mathsf x=
    \frac{\lVert \mathsf H \rVert^2}{\lambda_{\rm max}\op N_{\rm flux}}\right)
  \\
  \frac{\sqrt{3}}{2} + c^2\op\frac{2}{\sqrt{3}}\leq &\: \mathsf y \leq 
  \frac{\lambda_{\rm max}\op N_{\rm flux}}{\lVert \mathsf H \rVert^2} + c^2 \op \frac{\lVert \mathsf H \rVert^2}{\lambda_{\rm max}\op N_{\rm flux}}
  \\
  \frac{\sqrt{3}}{2} \leq &\: \mathsf y \leq 
  \frac{\lambda_{\rm max}\op N_{\rm flux}}{\lVert \mathsf H \rVert^2} + \frac{1}{4} \op \frac{\lVert \mathsf H \rVert^2}{\lambda_{\rm max}\op N_{\rm flux}}\,.
}  
Expressing then 
$  \int F\wedge\star F$ in matrix notation and using bounds similar to \eqref{rel_102},
we can write \eqref{rel_2051} as
\eq{
  \label{rel_206}
  \frac{\sqrt{3}}{2}\op \frac{N_{\rm flux}}{\lambda_{\rm max}} \leq \lVert \mathsf F \rVert^2
  \leq 
  \frac{\lambda^2_{\rm max}\op N_{\rm flux}^2}{\lVert \mathsf H \rVert^2} + \frac{1}{4} \op\lVert \mathsf H \rVert^2\,.
}
These inequalities should be understood for a given $H$-flux, that means, for a given flux-vector $\mathsf H$ the allowed choices for $\mathsf F$ are bounded from above and below as shown in \eqref{rel_206}.
Especially for large $\lVert \mathsf H\rVert^2$, these bounds are typically stronger than 
\eqref{rel_210}.

%%%%%%%%%%%%%%%%%%%%%%%%%%%%%%%%%%%%%%%%%%%%%%%
%%%%%%%%%%%%%%%%%%%%%%%%%%%%%%%%%%%%%%%%%%%%%%%

\subsubsection*{Remark}

For regions with large maximal eigenvalues $\lambda_{\rm max}$ of the Hodge-star 
matrix $\mathcal M$, the number of flux choices within the above bounds can be 
extremely large. 
A possible way to restrict the flux choices is to express $\mathsf H^T\! \mathcal M\op  \mathsf H$
(and similarly $\mathsf F^T\! \mathcal M\op  \mathsf F$) as follows. 
We first recall from \eqref{rel_270} the splitting of the $H$-flux into two $(h^{2,1}_-+1)$-dimensional 
vectors $\mathsf h_1=h^I$ and $\mathsf h_2=h_I$. 
Using the explicit form of \eqref{mat_197} we can then write
\eq{
  \label{rel_394875957}
  \mathsf H^T\! \mathcal M\op  \mathsf H
  = \mathsf h_1^T (-{\mathcal I})\, \mathsf h_1
  + (\mathsf h_2 + \mathcal R\op \mathsf h_1)^T (-{\mathcal I})^{-1} 
   (\mathsf h_2 + \mathcal R\op \mathsf h_1)   
   \,.
}
Since $-\mathcal I$ is required to be positive definite, \eqref{rel_394875957}
is a sum of two semi-positive terms. We therefore have the bound
\eq{
\label{bound_123}
  \mu_{\rm min} \lVert \mathsf h_1 \rVert^2 \leq\mathsf h_1^T (-{\mathcal I})\, \mathsf h_1
  \leq  \mathsf H^T\! \mathcal M\op  \mathsf H\,,
}
where $\mu_{\rm min}$ is the smallest eigenvalue of $-\mathcal I$
and $\lVert \mathsf h_1 \rVert^2=\mathsf h_1^T\mathsf h_1$. This bound can 
be helpful to determine flux choices more efficiently.

%%%%%%%%%%%%%%%%%%%%%%%%%%%%%%%%%%%%%%%%%%%%%%%
%%%%%%%%%%%%%%%%%%%%%%%%%%%%%%%%%%%%%%%%%%%%%%%
%%%%%%%%%%%%%%%%%%%%%%%%%%%%%%%%%%%%%%%%%%%%%%%
%%%%%%%%%%%%%%%%%%%%%%%%%%%%%%%%%%%%%%%%%%%%%%%
%%%%%%%%%%%%%%%%%%%%%%%%%%%%%%%%%%%%%%%%%%%%%%%
%%%%%%%%%%%%%%%%%%%%%%%%%%%%%%%%%%%%%%%%%%%%%%%
%%%%%%%%%%%%%%%%%%%%%%%%%%%%%%%%%%%%%%%%%%%%%%%
%%%%%%%%%%%%%%%%%%%%%%%%%%%%%%%%%%%%%%%%%%%%%%%
%%%%%%%%%%%%%%%%%%%%%%%%%%%%%%%%%%%%%%%%%%%%%%%
%%%%%%%%%%%%%%%%%%%%%%%%%%%%%%%%%%%%%%%%%%%%%%%

\clearpage
\section{The mirror octic}
\label{sec_octic}

In this section we describe the compactification space we are considering for the following.
Our model is an orientifold projection of the mirror-octic three-fold with  $h^{2,1}_-=1$
and, hence, we study moduli stabilization for 
the axio-dilaton $\tau$ and one complex-structure modulus $z$.

%%%%%%%%%%%%%%%%%%%%%%%%%%%%%%%%%%%%%%%%%%%%%%%
%%%%%%%%%%%%%%%%%%%%%%%%%%%%%%%%%%%%%%%%%%%%%%%

\subsubsection*{Motivation}

In this paper we consider 
the mirror-dual of the octic hypersurface $X_8$ in weighted projective space $\mathbb{P}_{11114}$. Our motivation for choosing this model is as follows:
\begin{itemize}

\item 
First, the octic is one of the $14$ one-parameter hypergeometric models, i.e. 
the periods $\Pi$ are expressible purely in terms of hypergeometric functions and their parameter derivatives. In this regard the octic is even more special, as it is one of only four models in which the periods around the Landau-Ginzburg (LG) point are expressible purely in terms of hypergeometric functions. 

\item Second, an explicit orientifold projection that leads to a D3-brane tadpole charge $Q_{\rm D3}=8$ 
and to $h^{2,1}_-=1$ exists, where the D7-branes are placed on top of the O7-planes.\footnote{ We thank Jakob Moritz for determining  orientifolds of the one-parameter models using the techniques described in \cite{Moritz:2023jdb} and sharing this information with us.
The precise form of the orientifold projection is not needed for our purposes.
} 
For the four purely hypergeometric models 
the D3-brane tadpole charge is typically much larger: 
for the mirror octic  another orientifold projection with $Q_{\rm D3}=972$ can be found in \cite{Giryavets:2003vd} 
and for the other three examples one obtains with the techniques of \cite{Moritz:2023jdb} the values $Q_{\rm D3}=106$, $Q_{\rm D3}=148$, and $Q_{\rm D3}=152$.\footnote{
Flux vacua of the mirror-quintic have been studied for instance in \cite{CaboBizet:2016uzv}.
} Such large numbers render a complete scan of all vacua with our methods infeasible and we will therefore focus on the study of the $Q_{\rm D3}=8$ orientifold of the mirror octic.

\end{itemize}

%%%%%%%%%%%%%%%%%%%%%%%%%%%%%%%%%%%%%%%%%%%%%%%
%%%%%%%%%%%%%%%%%%%%%%%%%%%%%%%%%%%%%%%%%%%%%%%

\subsubsection*{The periods}

Since the octic is one of the 14 hypergeometric models its periods have been studied in a number of papers
\cite{Morrison:1991cd,Klemm:1992tx,Font:1992uk}. A modern summary of the necessary techniques can be found in \cite{Bastian:2023shf}, to which we refer for technical details.
The model has the topological data
\eq{
    \arraycolsep2pt
    \begin{array}{@{}lcr@{\hspace{25pt}}lcr@{}}
    h^{2,1}_-&=&1\,, & h^{2,1}_+&=&0\,,
    \\[6pt]
    h^{1,1}_-&=&72\,, & h^{1,1}_+&=&77\,,
    \end{array}
    \hspace{25pt}
    \chi=-296\,, \hspace{25pt}
    \kappa=2\,, \hspace{25pt}
    \int c_2\wedge J=44\,,
}
where $\chi$ denotes the Euler number, $\kappa$ denotes the 
triple-intersection number, $c_2$ is the second Chern class, and $J$ 
is the $(1,1)$-form on the mirror side. 
The periods $\Pi$ of this model follow  from the Picard-Fuchs equation
\begin{equation}
  \label{PF}
   \Bigl[ \theta^4-z (\theta+\tfrac{1}{8})(\theta+\tfrac{3}{8})(\theta+\tfrac{5}{8})(\theta+\tfrac{7}{8})\Bigr]\Pi=0\,,
\end{equation}
where $z$ is the complex-structure modulus and where $\theta=z\frac{\partial}{\partial z}$. 
The solutions to \eqref{PF} are typically expanded in power series that converge absolutely for $|z|<1$ and have to be analytically continued for $|z|\ge 1$.
As will be explained below, for us it is necessary to solve \eqref{PF} not only in the 
vicinity of $z=0$ but also around different expansion points. 
Let us therefore 
introduce a new variable $\tilde z$ and the corresponding differential operator $\tilde \theta$ as
\eq{
  z = z_{\rm ep} + \tilde z\,,
  \hspace{50pt}
  \tilde{\theta}=( z_{\rm ep} + \tilde z)\op\frac{\partial}{\partial \tilde z}\,,
}
where $z_{\rm ep}$ denotes the expansion point. In this variable  the differential equation \eqref{PF} 
takes the form
\begin{equation}
   \Bigl[ \tilde{\theta}^4-( z_{\rm ep} + \tilde z) (\tilde{\theta}+\tfrac{1}{8})(\tilde{\theta}+\tfrac{3}{8})(\tilde{\theta}+\tfrac{5}{8})(\tilde{\theta}+\tfrac{7}{8})\Bigr]\Pi=0\,.
\end{equation}
For the region around $z=\infty$ we introduce the variable $z_{\infty}$ and the corresponding differential operator 
$\theta_{\infty}$ as
\eq{
  \label{coords_01}
  z_\infty=\frac{1}{z}\,, 
  \hspace{50pt}
  \theta_{\infty}=z_{\infty}\op\frac{\partial}{\partial z_{\infty}}\,.
}
Here
 the Picard-Fuchs  equation \eqref{PF} becomes
\begin{equation}
   \Bigl[ z_\infty \theta_{\infty}^4- (\theta_{\infty}-\tfrac{1}{8})(\theta_{\infty}-\tfrac{3}{8})(\theta_{\infty}-\tfrac{5}{8})(\theta_{\infty}-\tfrac{7}{8})\Bigr]\Pi=0\,.
\end{equation}

%%%%%%%%%%%%%%%%%%%%%%%%%%%%%%%%%%%%%%%%%%%%%%%
%%%%%%%%%%%%%%%%%%%%%%%%%%%%%%%%%%%%%%%%%%%%%%%

\subsubsection*{Expansion points}

In theory the expansions around  $z=0$ and $z=\infty$ cover the whole moduli space. In practice, as one can evaluate the expansion only up to a finite order in $z$ and $z_{\infty}$, one can trust them
 only in a smaller region. 
To be able to  determine flux vacua for the whole moduli space, we need a global expression for the period vector $\Pi$. To achieve this, the periods are expanded around the following  points:
\begin{itemize}

\item the large-complex-structure (LCS) point at $z=0$, 

\item the conifold point at $z=+1$, 

\item the Landau-Ginzburg (LG) point at $z=\infty$,

\item the points $z=-1$ and $z=\pm i$. 

\end{itemize}
In total we therefore use expansions around six points, which  together cover all of the moduli space.

%%%%%%%%%%%%%%%%%%%%%%%%%%%%%%%%%%%%%%%%%%%%%%%
%%%%%%%%%%%%%%%%%%%%%%%%%%%%%%%%%%%%%%%%%%%%%%%

\subsubsection*{Visualizing the moduli space}

\begin{figure}[t]
\centering
\includegraphics[width=200pt]{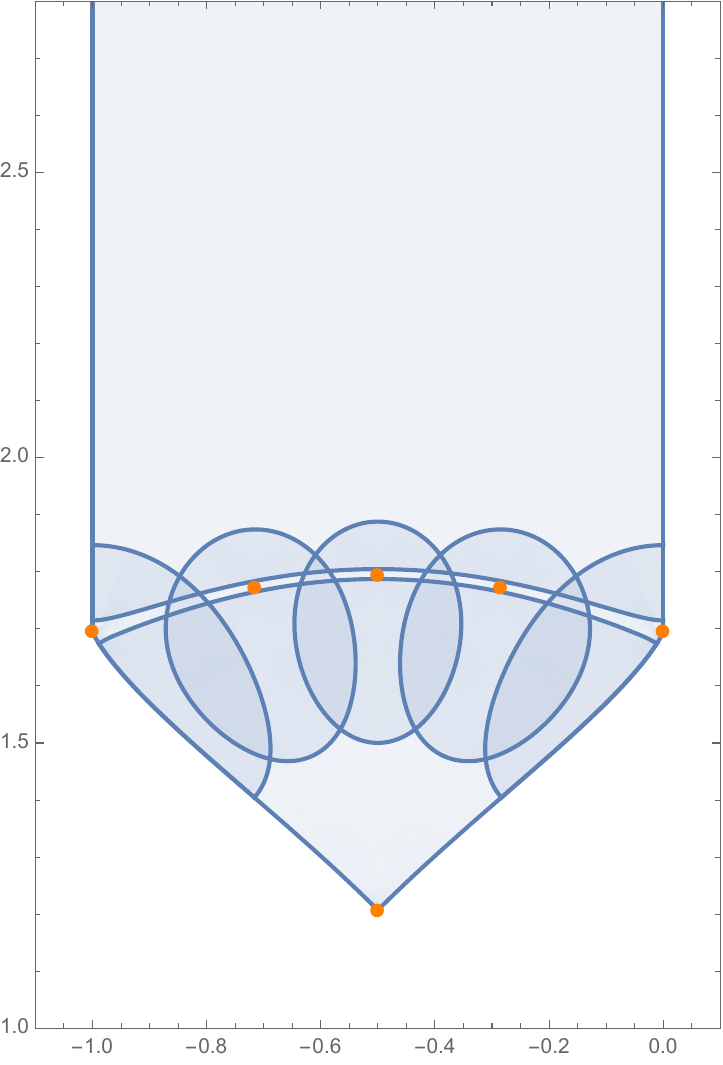}%
\begin{picture}(0,0)
\put(-96,185){\vector(0,-1){45}}
\put(-108,190){\scriptsize$z=-1$}
\put(-62,170){\vector(0,-1){32}}
\put(-74,175){\scriptsize$z=-i$}
\put(-130,170){\vector(0,-1){32}}
\put(-142,175){\scriptsize$z=+i$}
\put(-96,250){\vector(0,+1){45}}
\put(-105,242){\scriptsize$z=0$}
\put(-58,43){\vector(-1,0){30}}
\put(-53,41){\scriptsize$z=\infty$}
\put(19,120){\vector(-1,0){30}}
\put(22,118){\scriptsize$z=+1$}
\put(-210,120){\vector(+1,0){30}}
\put(-240,118){\scriptsize$z=+1$}
\put(-18,-7){\scriptsize$\mbox{Re}(t)$}
\put(-215,287){\scriptsize$\mbox{Im}(t)$}
\end{picture}
\vskip10pt
\caption{Covering of the fundamental domain in K\"ahler coordinates with the six bases described in the main text. The point $z=0$ corresponds to the LCS point, 
$z=1$ is the conifold, and $z=\infty$ is the LG point.
The overlapping regions correspond to discs of radius $0.9$ around the expansion points on the complex-structure side.
\label{FundamentalDomainCover}
}
\end{figure}

To visualize the global moduli space, it is useful to map the
complex-structure moduli space to the K\"ahler moduli space of the mirror manifold. This mapping is achieved by the mirror map $t(z)$, which in the integral symplectic basis is given by 
(c.f.~the expansion of $\Omega$ shown in \eqref{omega})
\begin{equation}
  \label{coord_kaehler}
    t(z)=\frac{\Pi^2(z)}{\Pi^1(z)}\;.
\end{equation}
In figure~\ref{FundamentalDomainCover} we have applied the mirror map and show the fundamental domain 
as well as the expansion points on the K\"ahler side. 
We have also mapped discs of radius $\lvert \tilde z\rvert =0.9$ on the complex-structure side centered around the expansion points to the K\"ahler side.
These regions overlap and allow us to cover the whole moduli space. 
Let us furthermore stress  that we view this purely as a change of coordinates to simplify visualization --- we will  stabilize the complex-structure modulus $z$ of the mirror octic and not the K\"ahler modulus $t$ of the mirror dual.

%%%%%%%%%%%%%%%%%%%%%%%%
%%%%%%%%%%%%%%%%%%%%%%%%
%%%%%%%%%%%%%%%%%%%%%%%%
%%%%%%%%%%%%%%%%%%%%%%%%

%%%%%%%%%%%%%%%%%%%%%%%%%%%%%%%%%%%%%%%%%%%%%%%
%%%%%%%%%%%%%%%%%%%%%%%%%%%%%%%%%%%%%%%%%%%%%%%

\subsubsection*{Remark}

The local expansions $\omega_i$ obtained by the Frobenius method
are  mapped by transition matrices $m_{\mathsf i}$ into the integral symplectic basis as
\begin{equation}
    \Pi=m_{\mathsf i}\cdot \omega_{\mathsf i}\,,
\end{equation}
where $\mathsf i$ labels the expansion point and
where $\omega_{\mathsf i}$ is a four-vector.
The computation of these matrices is technical and we refer to \cite{Bastian:2023shf} for  details. Here we just point out that for this specific model there exists the LG basis
\eq{
    \omega_{\rm{LG}}=z_{\infty}^{j/8}\;_5F_4\left( \left\{\frac{j}{8},\frac{j}{8},\frac{j}{8},\frac{j}{8},1\right\},
    \left\{\frac{j+1}{8},\frac{j+3}{8},\frac{j+5}{8},\frac{j+7}{8}
    \right\},z_{\infty}\right),
}
where the index $j=\{1,3,5,7\}$ corresponds to the $\frac{j+1}{2}$ entry of the four-vector $\omega_{\rm LG}$. Note that this basis consists of $\vphantom{F}_4F_3$ functions and the $\vphantom{F}_5F_4$ is just used to shorten the notation, as the $1$ will always cancel against one of the
lower parameters. This basis is very useful as the analytic continuation to the LCS point can be performed analytically using the connection formula of the $\vphantom{F}_4F_3$ functions and it can be evaluated numerically exact at all other expansion points\footnote{There is a technical issue at the conifold point where this basis diverges. For the conifold one has to evaluate the expansion around a point $z^{-1}=1-\epsilon$ for $\epsilon\ll 1$ that is slightly away from the conifold to ensure convergence.}, simplifying the computation of the transition matrices.

%%%%%%%%%%%%%%%%%%%%%%%%%%%%%%%%%%%%%%%%%%%%%%%
%%%%%%%%%%%%%%%%%%%%%%%%%%%%%%%%%%%%%%%%%%%%%%%
%%%%%%%%%%%%%%%%%%%%%%%%%%%%%%%%%%%%%%%%%%%%%%%
%%%%%%%%%%%%%%%%%%%%%%%%%%%%%%%%%%%%%%%%%%%%%%%
%%%%%%%%%%%%%%%%%%%%%%%%%%%%%%%%%%%%%%%%%%%%%%%
%%%%%%%%%%%%%%%%%%%%%%%%%%%%%%%%%%%%%%%%%%%%%%%
%%%%%%%%%%%%%%%%%%%%%%%%%%%%%%%%%%%%%%%%%%%%%%%
%%%%%%%%%%%%%%%%%%%%%%%%%%%%%%%%%%%%%%%%%%%%%%%
%%%%%%%%%%%%%%%%%%%%%%%%%%%%%%%%%%%%%%%%%%%%%%%
%%%%%%%%%%%%%%%%%%%%%%%%%%%%%%%%%%%%%%%%%%%%%%%

\clearpage
\section{All flux vacua}
\label{sec_strategy}

We now describe our strategy for  constructing 
all flux vacua corresponding to vanishing F-terms 
(c.f.~equation \eqref{min_99})
for the mirror-octic three-fold.
This section contains the technical details of 
our analysis --- the reader not interested in those
can skip to  section~\ref{sec_results} for a discussion of 
our results.

%%%%%%%%%%%%%%%%%%%%%%%%%%%%%%%%%%%%%%%%%%%%%%%
%%%%%%%%%%%%%%%%%%%%%%%%%%%%%%%%%%%%%%%%%%%%%%%
%%%%%%%%%%%%%%%%%%%%%%%%%%%%%%%%%%%%%%%%%%%%%%%
%%%%%%%%%%%%%%%%%%%%%%%%%%%%%%%%%%%%%%%%%%%%%%%
%%%%%%%%%%%%%%%%%%%%%%%%%%%%%%%%%%%%%%%%%%%%%%%

\subsection{General strategy}

Our general strategy to construct all 
flux vacua with vanishing F-terms \eqref{min_99} for the mirror-octic 
consists of  two parts.

%%%%%%%%%%%%%%%%%%%%%%%%%%%%%%%%%%%%%%%%%%%%%%%
%%%%%%%%%%%%%%%%%%%%%%%%%%%%%%%%%%%%%%%%%%%%%%%

\subsubsection*{Bulk regions}

\begin{figure}[t]
\centering
\includegraphics[width=225pt]{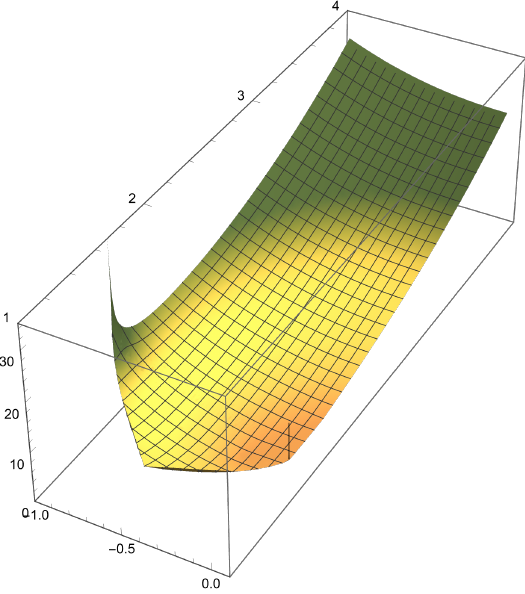}%
\begin{picture}(0,0)
\put(-148,-9){\scriptsize$\mbox{Re}(t)$}
\put(-106,248){\scriptsize$\mbox{Im}(t)$}
\put(-248,95){\scriptsize$\lambda_{\rm max}$}
\put(-248,95){\scriptsize$\lambda_{\rm max}$}
\put(-62,57){\vector(-1,0){35}}
\put(-58,55.5){\scriptsize conifold}
\put(-218,140){\vector(1,0){35}}
\put(-250,138){\scriptsize conifold}
\end{picture}
\vskip10pt
\caption{Dependence of the maximal eigenvalue $\lambda_{\rm max}$
of the Hodge-star matrix $\mathcal M$ on the complex-structure modulus,
where the plot-region corresponds to the moduli space on the K\"ahler side,
c.f.~figure~\ref{FundamentalDomainCover}. Towards the conifold 
and large-complex-structure points $\lambda_{\rm max}$ diverges, 
while $\lambda_{\rm max}$ stays finite in the bulk of the moduli space.
Note that the divergence for the conifold at $\mbox{Re}(t)=0$ is rather sharp and difficult to visualize.
\label{fig_hodge_eigenvalue}
}
\end{figure}

We first choose  regions 
in complex-structure moduli space in which the 
eigenvalues of the Hodge-star matrix
$\mathcal M$ are finite.
These regions have been shown in figure~\ref{FundamentalDomainCover} and
are characterized as follows 
\eq{
\label{regions_finite}
\arraycolsep12pt
\renewcommand{\arraystretch}{1.3}
\begin{array}{l@{\hspace{2pt}}c@{\hspace{2pt}}c c || l@{\hspace{2pt}}c@{\hspace{2pt}}l@{\hspace{2pt}}c@{\hspace{2pt}}l || r}
\multicolumn{4}{c||}{\mbox{expansion point}}
&
\multicolumn{5}{c||}{\mbox{region}}
&
\lambda_{\rm max}
\\
\hline\hline
z_{\rm ep}&=& 0 & \mbox{LCS} & 0.01 &\leq &-\frac{\log\lvert z-z_{\rm ep}\rvert}{2\pi} &\leq & 3 & 37.98
\\
z_{\rm ep}&=& +1 & \mbox{conifold} & 0.01 &\leq &-\frac{\log\lvert z-z_{\rm ep}\rvert}{2\pi} &\leq & 3 & 3.95
\\
z_{\rm ep}&=& \infty & \mbox{LG} & 0.01 &\leq &+\frac{\log\lvert z_{\infty}\rvert}{2\pi} & & & 8.33
\\
z_{\rm ep}&=& -1 &  & 0.01 &\leq &-\frac{\log\lvert z-z_{\rm ep}\rvert}{2\pi} & & & 5.00
\\
z_{\rm ep}&=& +i &  & 0.01 &\leq &-\frac{\log\lvert z-z_{\rm ep}\rvert}{2\pi} & & & 11.24
\\
z_{\rm ep}&=& -i &  & 0.01 &\leq &-\frac{\log\lvert z-z_{\rm ep}\rvert}{2\pi} & & & 3.35
\end{array}
}
where $z_{\infty}=1/z$ for the Landau-Ginzburg point was introduced in \eqref{coords_01}.
Here $\lambda_{\rm max}$ denotes the maximal eigenvalue of $\mathcal M$ within the
specified region, and 
the dependence of $\lambda_{\rm max}$ on the whole complex-structure moduli 
space is shown in figure~\ref{fig_hodge_eigenvalue}.\footnote{
Although the Hodge-star matrix $\mathcal M$, in particular its maximal 
eigenvalue $\lambda_{\rm max}$, is not symmetric about the $\mbox{Re}(t)=-\frac{1}{2}$ axis,
the flux vacua are distributed symmetrically about this axis.
}
Using the bounds on the fluxes derived in section~\ref{sec_bounds}, we
can then construct a finite set of fluxes for each region. 
For these fluxes we solve the minimum conditions \eqref{rel_401}
and keep only those flux combinations that give rise 
to solutions within the chosen search region in complex-structure moduli space
and within the fundamental domain \eqref{tau_fundamental_d} of the axio-dilaton.

%%%%%%%%%%%%%%%%%%%%%%%%
%%%%%%%%%%%%%%%%%%%%%%%%
%%%%%%%%%%%%%%%%%%%%%%%%
%%%%%%%%%%%%%%%%%%%%%%%%

%%%%%%%%%%%%%%%%%%%%%%%%%%%%%%%%%%%%%%%%%%%%%%%
%%%%%%%%%%%%%%%%%%%%%%%%%%%%%%%%%%%%%%%%%%%%%%%

\subsubsection*{Boundary regions}

However, the  regions in \eqref{regions_finite}
do not  cover the moduli space
of the mirror octic completely. In particular, for the regions around the conifold and LCS point
the maximal eigenvalue of $\mathcal M$ diverges. We therefore also consider the regions
\eq{
\label{regions_sing}
\arraycolsep12pt
\renewcommand{\arraystretch}{1.3}
\begin{array}{l@{\hspace{2pt}}c@{\hspace{2pt}}c c || l@{\hspace{2pt}}c@{\hspace{2pt}}l }
\multicolumn{4}{c||}{\mbox{expansion point}}
&
\multicolumn{3}{c}{\mbox{region}}
\\
\hline\hline
z_{\rm ep}&=& 0 & \mbox{LCS} & 3 &\leq &-\frac{\log\lvert z-z_{\rm ep}\rvert}{2\pi} 
\\
z_{\rm ep}&=& +1 & \mbox{conifold} & 3 &\leq &-\frac{\log\lvert z-z_{\rm ep}\rvert}{2\pi} 
\end{array}
}
for which we can  ignore higher-order corrections to the periods
and only work with the lowest-order expressions. 
As will be explained in detail in sections
\ref{sec_region_lcs}
and \ref{sec_region_coni},
for these cases we can classify all possible flux choices and determine 
the flux vacua.

%%%%%%%%%%%%%%%%%%%%%%%%%%%%%%%%%%%%%%%%%%%%%%%
%%%%%%%%%%%%%%%%%%%%%%%%%%%%%%%%%%%%%%%%%%%%%%%
%%%%%%%%%%%%%%%%%%%%%%%%%%%%%%%%%%%%%%%%%%%%%%%
%%%%%%%%%%%%%%%%%%%%%%%%%%%%%%%%%%%%%%%%%%%%%%%
%%%%%%%%%%%%%%%%%%%%%%%%%%%%%%%%%%%%%%%%%%%%%%%

\subsection{Finite regions}
\label{sec_finite_regions}

We now describe the technical details 
of constructing the flux vacua in the finite regions
of moduli space. 
For all the regions
defined in \eqref{regions_finite} we essentially perform the 
following steps:
\begin{enumerate}

\item We  express the complex-structure modulus $z$ 
in terms of real variables $u$ and $v$ 
as 
\eq{
\label{coord_exp}
 z-z_{\rm ep}=\tilde z =e^{2\pi\op i(u+i\op v)}\,,
} 
where $z_{\rm ep}$ denotes the expansion point, 
$u\in(-\frac12,+\frac12]$ and $v\geq 0.01$. 
For the LG point we employ the expansion
$z^{-1}=z_{\infty} = e^{2\pi\op i(u+i\op v)}$.

\item 
In each region we construct a grid $\mathsf G(u,v)$ in the variables $u$ and $v$, 
and for each point of the grid we compute the matrices $\mathcal M$ and $\mathcal I$
at order $50$ in $e^{-2\pi\op v}$, that is, we expand the periods up to order $50$ in $\tilde z$.
In  figure~\ref{fig_grid} we show the maximal eigenvalue $\lambda_{\rm max}$ of $\mathcal M$
on the grid for the conifold region.
%%%%%%%%%%%%%%%%%%%%%%%%
%%%%%%%%%%%%%%%%%%%%%%%%
\begin{figure}[t]
\hspace{\stretch{1}}
\includegraphics[width=280pt]{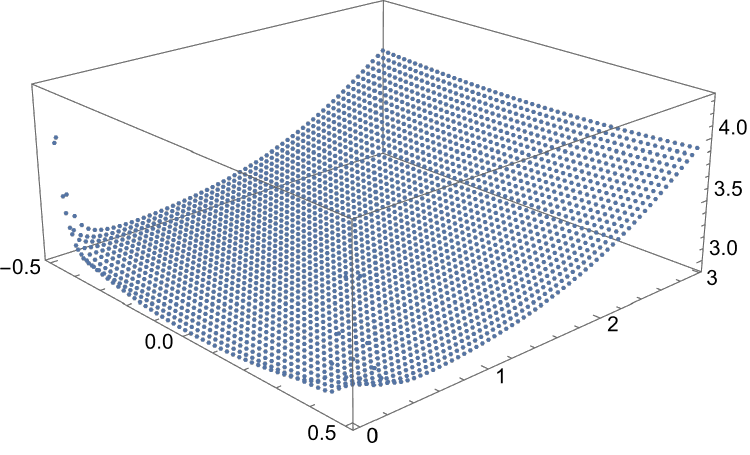}
\begin{picture}(0,0)
\put(-270,55){\footnotesize $u$}
\put(-28,49){\footnotesize $v$}
\put(-2,128){\footnotesize $\lambda_{\rm max}$}
\end{picture}
\hspace{\stretch{1}}
\vspace*{10pt}
\caption{Plot of the maximal eigenvalue $\lambda_{\rm max}$ 
of the Hodge-star matrix $\mathcal M$ on a grid $\mathsf G(u,v)$.
The region shown here is around the conifold point $z_{\rm ep}=1$.
The divergence of  $\lambda_{\rm max}$ at $u=\pm 1/2$ for $v\to 0$
corresponds to the LCS point.
\label{fig_grid}}
\end{figure}
%%%%%%%%%%%%%%%%%%%%%%%%
%%%%%%%%%%%%%%%%%%%%%%%%

\item We then construct all four-dimensional flux vectors $\mathsf H$ that satisfy the bound 
\eqref{rel_121}. When necessary, we 
first employ the bound \eqref{bound_123}.

\item We compute $\mathsf H^T\! \mathcal M\op  \mathsf H$ 
for each point of the grid $\mathsf G(u,v)$. If the relation
\eq{
  \mathsf H^T\! \mathcal M\op  \mathsf H \leq \frac{2\op N_{\rm max}}{\sqrt{3}}
}
can   approximately be satisfied for at least one point of the grid, we keep the flux choice
together with the minimal value $ (\mathsf H^T\! \mathcal M\op  \mathsf H)_{\rm min}$
of $ \mathsf H^T\! \mathcal M\op  \mathsf H$ on the grid
$\mathsf G(u,v)$.

\item We construct all flux vectors $\mathsf F$ that satisfy an improved  version of the
bound \eqref{rel_206}. In particular, using
$ (\mathsf H^T\! \mathcal M\op  \mathsf H)_{\rm min}$ instead of $\lVert \mathsf H \rVert^2$ we solve 
for each choice of $\mathsf H$ 
the following relation for $\mathsf F$
\eq{
\frac{1}{\lambda_{\rm max}}\lVert \mathsf F \rVert^2
  \leq 
  \frac{N_{\rm flux}^2}{(\mathsf H^T\! \mathcal M\op  \mathsf H)_{\rm min}} 
  + \frac{1}{4} \op(\mathsf H^T\! \mathcal M\op  \mathsf H)_{\rm min}\,.
}

\item The minimum condition can be expressed in matrix form 
as shown in \eqref{min_204}. For each flux choice $(\mathsf H,\mathsf F)$ we then evaluate the vector 
\eq{
  \label{min_vec}
  \mathsf F - \bigl( \mathds 1\op c + \eta \op \mathcal M \op s\bigr) \mathsf H
}
for each point of the grid $\mathsf G(u,v)$. 
If there exists at least one point on the grid for which the norm of the vector \eqref{min_vec} is smaller than some cutoff,  we keep the flux combination.

\item Using the matrix version of \eqref{min_102} we determine the 
axio-dilaton on each point of the grid for the flux choices $(\mathsf H,\mathsf F)$.
If there exists at least one point in the grid for which 
 the axio-dilaton is stabilized approximately within the fundamental domain \eqref{tau_fundamental_d},
 we keep the flux choice.

\end{enumerate}
The steps described above ensure that we construct all flux choices 
relevant for the chosen region. The filtering was necessary in order
to keep the number of flux choices manageable, especially for large 
values of $N_{\rm flux}$.
The next step is to search for minima:
\begin{enumerate}[resume]

\item We numerically solve the condition \eqref{rel_401} 
for each remaining flux choice $(\mathsf H,\mathsf F)$ in the corresponding region. We use
\texttt{Mathematica}'s
\texttt{NMinimize} routine and  specify as 
method a random search with $30$ search points.
These computations are done at order $100$ in the expansion of the periods.

\item The minima obtained in the previous step are 
checked at order $200$ with 
\texttt{Mathematica}'s 
\texttt{FindMinimum} routine.

\end{enumerate}

%%%%%%%%%%%%%%%%%%%%%%%%%%%%%%%%%%%%%%%%%%%%%%%
%%%%%%%%%%%%%%%%%%%%%%%%%%%%%%%%%%%%%%%%%%%%%%%
%%%%%%%%%%%%%%%%%%%%%%%%%%%%%%%%%%%%%%%%%%%%%%%
%%%%%%%%%%%%%%%%%%%%%%%%%%%%%%%%%%%%%%%%%%%%%%%
%%%%%%%%%%%%%%%%%%%%%%%%%%%%%%%%%%%%%%%%%%%%%%%

\subsection{The LCS point}
\label{sec_region_lcs}

We now discuss the region in complex-structure moduli space
around the LCS point $z=0$. Since at this point 
the maximal eigenvalue of the Hodge-star matrix $\mathcal M$
diverges, the approach outlined above is in general not applicable.

%%%%%%%%%%%%%%%%%%%%%%%%%%%%%%%%%%%%%%%%%%%%%%%
%%%%%%%%%%%%%%%%%%%%%%%%%%%%%%%%%%%%%%%%%%%%%%%

\subsubsection*{Fluxes with $\mathsf h_1\neq 0$}

Let us first recall  our discussion around equation 
\eqref{rel_394875957} where we  split the 
$\mathsf H$-flux into the  
vectors $\mathsf h_1=(h^0,h^1)$ and $\mathsf h_2=(h_0,h_1)$. 
In this paragraph we consider the case $\mathsf h_1\neq0$, 
for which we find from \eqref{bound_123}
\eq{
  \label{rel_3001}
  \mu_{\rm min} \leq \mu_{\rm min} \lVert \mathsf h_1 \rVert^2 
  \leq  \mathsf H^T\! \mathcal M\op  \mathsf H\,,
}
where $\mu_{\rm min}\equiv \mu_{\rm min}(u,v)$ is the smallest eigenvalue of $-\mathcal I(u,v)$. Using then the bounds \eqref{rel_3001} and \eqref{rel_202} 
and the solution for the dilaton \eqref{min_102},
we obtain
\eq{
  \label{rel_3002}
  \mu_{\rm min} \leq \frac{2\op N_{\rm max}}{\sqrt{3}} \,. 
}
For the LCS point of the mirror octic it  turns out that 
$\mu_{\rm min}$ is monotonically-increasing for $v\to \infty$, 
and hence there is a $v_{\rm max}$ until which 
\eqref{rel_3002} is satisfied.
This implies that for $\mathsf h_1\neq0$ the relevant region 
around the LCS point is finite and we can employ the 
techniques explained in section~\ref{sec_finite_regions}.

%%%%%%%%%%%%%%%%%%%%%%%%%%%%%%%%%%%%%%%%%%%%%%%
%%%%%%%%%%%%%%%%%%%%%%%%%%%%%%%%%%%%%%%%%%%%%%%

\subsubsection*{Fluxes with $\mathsf h_1= 0$}

For flux choices with $\mathsf h_1=0$ 
we consider the near-boundary region and ignore 
exponential (i.e.~instanton) corrections to the period vector --- we keep however all polynomial 
terms in $u$ and $v$ including the $\alpha'$-corrections.
In this case the 
minimum conditions \eqref{min_204} 
are simplified.
For notational convenience we now introduce
\eq{
  \hat u = u\,, \hspace{40pt}
  \hat v = v + \frac{16\log 2}{2\pi}\,,
  \hspace{60pt}
  \xi=\frac{111\op\zeta(3)}{(2\pi)^3\op \hat v^3}\,,
}
where the shift in $v$ corresponds to the number $\mu$ appearing for instance 
in table~1 of \cite{Bastian:2023shf}.
In this notation one of the two independent minimum conditions contained in \eqref{min_204} 
can be expressed as follows
\eq{
 \label{rel_2009}
 (1-4\op\xi)\, \hat v^2 \op f^0\op h_1 = 3 \,(f^1-\hat u\op f^0)(h_0+\hat u\op h_1) \,.
}
We then distinguish the following three cases:
\begin{itemize}

\item For $f^0\neq 0$ and $h_1\neq 0$ we note that the flux number 
can be written as
\eq{
N_{\rm flux} = f^0\op  (h_0+\hat u \op h_1) +  (f^1 - \hat u\op f^0) \op h_1 \,.
}
Solving this relation for say $f^0$ and using it in \eqref{rel_2009}, 
we can derive the bound
\eq{
 (1-4\op\xi)\, \hat v^2 \leq \frac{3}{4}\op N_{\rm flux}^2 \,.
}
For a finite value of $N_{\rm flux}$ this implies that $v$ is bounded from 
above, and hence the region in moduli space we need to consider is finite. 
We can then apply the techniques 
explained in section~\ref{sec_finite_regions}
for constructing a finite set of flux choices.

\item For $f^0= 0$ we need to require $h_1\neq0$ in order to obtain 
a non-vanishing flux number. 
In particular, we have $N_{\rm flux} = f^1 h_1$
and for finite $N_{\rm flux}$ the allowed values for $h_1$ 
 and $f^1$ are finite.
The condition \eqref{rel_2009}
is then solved as
\eq{
  \hat u = - \frac{h_0}{h_1} \,,
}
and requiring $-\frac{1}{2}< \hat u \leq +\frac12$
leads to finitely-many choices for $h_0$.
The remaining minimum condition takes the form
\eq{
  \label{rel_6001}
  -\frac{18\op \xi}{1-\xi}\, N_{\rm flux}\op \hat v^2 = 
  6 \left( f_1 \op h_0 - f_0 \op h_1\right)+\frac{f^1}{h_1} \left(
  6\op  h_0^2 - 11\op  h_1^2 \right),
}
where  only $f_0$ and $f_1$ are unspecified.
These can be restricted by solving \eqref{rel_6001} for $\hat v$,
using this solution to determine the axio-dilaton 
 $\tau=c+i\op s$, and requiring $\tau$  to be in the fundamental domain.
In this way all flux choices can be constructed.

\item For $h_1=0$ we proceed along similar lines. 
The flux number takes the form $N_{\rm flux} = f^0 h_0$,
and for finite $N_{\rm flux}$ there are only finitely-many choices
for $f^0$ and $h_0$. 
The condition \eqref{rel_2009} is solved by
\eq{
  \hat u = + \frac{f^1}{f^0} \,,
}
and requiring  $-\frac{1}{2}< \hat u \leq +\frac12$
fixes the range of $f^1$.
The remaining minimum condition takes the form
\eq{
  \label{rel_6002}
\frac{18\op \xi}{1-\xi}\, \hat v^2 = 11+6\left(\frac{f_1}{f^0}-
\left[ \frac{f^1}{f^0} \right]^2 \right),
}
where the flux quanta $f_0$ and  $f_1$ are so far not restricted.
Solving \eqref{rel_6002} for $\hat v$, using this solution to determine 
$\tau$, and  requiring the axio-dilaton to be stabilized in the fundamental domain
leads to a finite number of choices for $f_0$ and  $f_1$.

\end{itemize}
The three cases discussed above cover all possible  flux choices for the situation $\mathsf h_1=0$. 
For each flux choice we then determine an approximate 
solution to the minimum condition \eqref{rel_401}, which is then 
checked at a higher order in the expansion of the periods. 
We finally note that the special cases $f^0=0$ and $h_1=0$ 
allow for large stabilized values of $v$ at a relatively low 
flux number.

%%%%%%%%%%%%%%%%%%%%%%%%%%%%%%%%%%%%%%%%%%%%%%%
%%%%%%%%%%%%%%%%%%%%%%%%%%%%%%%%%%%%%%%%%%%%%%%
%%%%%%%%%%%%%%%%%%%%%%%%%%%%%%%%%%%%%%%%%%%%%%%
%%%%%%%%%%%%%%%%%%%%%%%%%%%%%%%%%%%%%%%%%%%%%%%
%%%%%%%%%%%%%%%%%%%%%%%%%%%%%%%%%%%%%%%%%%%%%%%

\subsection{The conifold point}
\label{sec_region_coni}

Next, we  turn to the second singular point 
in  complex-structure moduli space, namely 
the conifold point. 
\begin{itemize}

\item 
We again introduce 
real variables $u$ and $v$ 
as $ z-1=\tilde z =e^{2\pi\op i(u+i\op v)}$, 
where $u\in(-\frac12,+\frac12]$
and $v\geq 3$.
Near the conifold point we determine the complex matrix 
\eqref{gkf} to leading order in $v$ as follows
\eq{
\label{rel_6012}
  \mathcal N = 
  \frac{1}{-u+i\op (v+\alpha)}
  \arraycolsep6pt
  \left( \begin{array}{cc}
  1 & i\op \beta \\[6pt]
   i\op\beta & -\beta^2 +\gamma\op(i\op u+v+\alpha)
  \end{array}
  \right),
  }
where the constants $\alpha$, $\beta$, $\gamma$ are given by
\eq{
  \alpha = 0.5728\ldots\,, \hspace{40pt}
  \beta = 0.5899\ldots\,, \hspace{40pt}
  \gamma = 2.7917\ldots\,.
}

\item With the help of \eqref{rel_6012} we  compute the Hodge-star matrix $\mathcal M$ from equation \eqref{mat_197}. For this $\mathcal M$ we determine
$\mathsf H^T\! \mathcal M\op  \mathsf H$, which can be arranged into a sum of positive 
terms as
\eq{
  \label{rel_6004}
  \mathsf H^T\! \mathcal M\op  \mathsf H
  =\hspace{16pt}&\frac{\gamma}{(v+\alpha)\op\gamma-\beta^2}  
  \left[ h^0- u\op h_1 + \frac{\beta}{\gamma}\op h_1\right]^2
  + \gamma\left[ h^1+ \tfrac{\beta}{\gamma}\op h_0\right]^2
  \\[8pt]
  +\,&\frac{(v+\alpha)\op\gamma-\beta^2}{\gamma} \, \bigl[ h_0 \bigr]^2
  + \frac{1}{\gamma} \, \bigl[ h_1 \bigr]^2\,.
}

\item In the approximation employed here, the minimum condition \eqref{min_204} can be solved 
analytically. From this solution we conclude the axio-dilaton can only be stabilized in its fundamental domain
if at least one of the $h^1, h_0,h_1$ and at least one of the $f^1, f_0,f_1$ are non-vanishing.

\item Using \eqref{rel_6004} and a corresponding expression for $\mathsf F^T\! \mathcal M\op  \mathsf F$
in the bounds \eqref{rel_121} and \eqref{rel_210}, respectively, 
all fluxes except $h^0$ and $f^0$ are restricted. 
The latter are fixed by requiring 
$-\frac{1}{2}< u \leq +\frac12$ and that the flux number $N_{\rm flux}$ is bounded.

\end{itemize}
In this way all flux vacua near the conifold point for a given maximal $N_{\rm flux}$ can be 
constructed.
The approximate solutions to the minimum condition \eqref{rel_401} can be determined
and are checked at order $200$ in the expansion of the periods. 

%%%%%%%%%%%%%%%%%%%%%%%%%%%%%%%%%%%%%%%%%%%%%%%
%%%%%%%%%%%%%%%%%%%%%%%%%%%%%%%%%%%%%%%%%%%%%%%
%%%%%%%%%%%%%%%%%%%%%%%%%%%%%%%%%%%%%%%%%%%%%%%
%%%%%%%%%%%%%%%%%%%%%%%%%%%%%%%%%%%%%%%%%%%%%%%
%%%%%%%%%%%%%%%%%%%%%%%%%%%%%%%%%%%%%%%%%%%%%%%

\subsection{The LG point}
\label{sec_region_LG}

In the region around the LG point $z=\infty$ the eigenvalues of the Hodge-star matrix 
$\mathcal M$ are finite and therefore the approach described in section~\ref{sec_finite_regions}
is applicable. 
However, we note that at the LG point the Hodge-star matrix becomes particularly simple, namely
\eq{
  \mathcal M_{\rm LG} = 
  \arraycolsep4pt
  \left( 
  \begin{array}{cccc}
  6 & 0 & -3 & 2 \\
  0 & 3 & 1 & 0 \\
  -3 & 1 & 2 & -1 \\
  2 & 0 & -1 & 1
  \end{array}
  \right).
  }
Since this matrix is integer-valued, bounds on the fluxes 
are more-easily implemented. Using \eqref{rel_202} and \eqref{rel_210111} we find
\eq{
  \label{bounds_1000}
  \mathsf H^T\! \mathcal M\op  \mathsf H \leq \frac{2\op N_{\rm flux}}{\sqrt{3}}\,,
  \hspace{70pt}
  \mathsf F^T \!\mathcal M\op \mathsf F \leq \frac{4}{3}\op \frac{N^2_{\rm flux}}{\mathsf H^T \!\mathcal M\op \mathsf H}\,,
}
and using the solutions \eqref{min_102} for the axio-dilaton in the minimum conditions \eqref{min_204}
we obtain
\eq{
  \label{bounds_1001}
  \eta \,\frac{ \mathsf H \op \mathsf F^T- \mathsf F\op \mathsf H^T}{N_{\rm flux}}\op \mathcal M \op \mathsf H 
  = \mathcal M \op \mathsf H\,.
}
The flux choices satisfying the bounds \eqref{bounds_1000} can then be inserted in \eqref{bounds_1001}
and be checked whether they correspond to a minimum. 
Since $  \mathcal M_{\rm LG}$ is integer-valued it also follows that the axio-dilaton $\tau = c+i\op s$ 
is stabilized at rational values.

%%%%%%%%%%%%%%%%%%%%%%%%%%%%%%%%%%%%%%%%%%%%%%%
%%%%%%%%%%%%%%%%%%%%%%%%%%%%%%%%%%%%%%%%%%%%%%%
%%%%%%%%%%%%%%%%%%%%%%%%%%%%%%%%%%%%%%%%%%%%%%%
%%%%%%%%%%%%%%%%%%%%%%%%%%%%%%%%%%%%%%%%%%%%%%%
%%%%%%%%%%%%%%%%%%%%%%%%%%%%%%%%%%%%%%%%%%%%%%%
%%%%%%%%%%%%%%%%%%%%%%%%%%%%%%%%%%%%%%%%%%%%%%%
%%%%%%%%%%%%%%%%%%%%%%%%%%%%%%%%%%%%%%%%%%%%%%%
%%%%%%%%%%%%%%%%%%%%%%%%%%%%%%%%%%%%%%%%%%%%%%%
%%%%%%%%%%%%%%%%%%%%%%%%%%%%%%%%%%%%%%%%%%%%%%%
%%%%%%%%%%%%%%%%%%%%%%%%%%%%%%%%%%%%%%%%%%%%%%%

\clearpage
\section{Results and discussion}
\label{sec_results}

In this section we present and discuss our results.
In this work we have chosen an orientifold projection for the mirror-octic three-fold that 
leads to a  D3-brane tadpole contribution of  $Q_{\rm D3}=8$
and hence, due to the tadpole cancellation condition \eqref{tadpole_d3},
the largest allowed flux number for this model is $N_{\rm flux}=8$.
However, in order to better understand the structure of the flux landscape, 
we  have determined all flux vacua 
for  $N_{\rm flux}\leq10$.

%%%%%%%%%%%%%%%%%%%%%%%%%%%%%%%%%%%%%%%%%%%%%%%
%%%%%%%%%%%%%%%%%%%%%%%%%%%%%%%%%%%%%%%%%%%%%%%
%%%%%%%%%%%%%%%%%%%%%%%%%%%%%%%%%%%%%%%%%%%%%%%
%%%%%%%%%%%%%%%%%%%%%%%%%%%%%%%%%%%%%%%%%%%%%%%
%%%%%%%%%%%%%%%%%%%%%%%%%%%%%%%%%%%%%%%%%%%%%%%

\subsection{Distribution of vacua}

\begin{figure}[p]
\vspace{30pt}
\hspace{\stretch{1}}
\includegraphics[width=120pt]{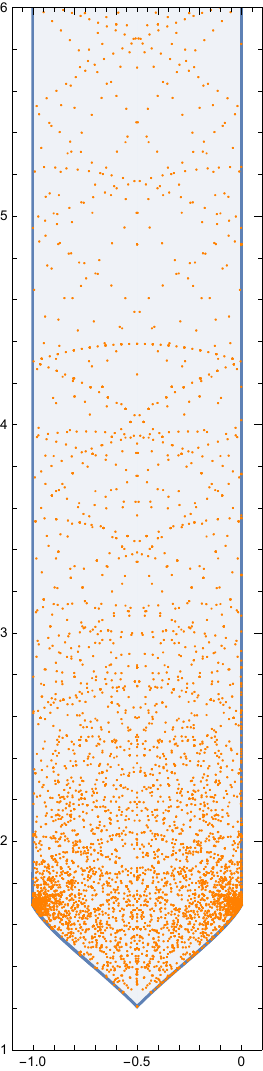}
\hspace{\stretch{1.04}}
\vspace*{10pt}
\caption{Distribution of 
stabilized complex-structure moduli on the K\"ahler side
for  values of the flux number $1\leq N_{\rm flux}\leq10$.
\label{fig_minima_all}}
\end{figure}

In figures~\ref{fig_minima} we have show 
the location of the stabilized complex-structure moduli
for each flux number $N_{\rm flux}$, and 
in figure~\ref{fig_minima_all} the combined distribution 
for all values $N_{\rm flux}=1,\ldots,10$ can be found.
The plots are for the region $-1\leq \mbox{Re}(t)\leq 0$ and 
$1\leq \mbox{Im}(t)\leq 6$
on the K\"ahler side but, especially for large $N_{\rm flux}$, 
there are vacua with $\mbox{Im}(t)> 6$ that are not shown.
We make the following qualitative observations:
\begin{itemize}

\item The distribution of points is symmetric with respect to the $\mbox{Re}(t)=-\frac{1}{2}$ axis.
However, the boundary on the left-hand side is identified with the right-hand side
and therefore boundary points do not respect this symmetry. 

\item Around the conifold point (c.f.~figure~\ref{FundamentalDomainCover})
we observe a higher density of vacua, in agreement with the arguments made in \cite{Denef:2004ze}.

\item Especially in the LCS regime arc-like structures can be 
identified, but we were not able to describe them analytically.

\item Vacua are found directly at the LG point, but there is a small 
void around the LG point in which no vacua are located. 

\item Vacua with small flux number  are located in the interior of moduli space, 
in agreement with arguments made in \cite{Plauschinn:2021hkp}.

\end{itemize}

%%%%%%%%%%%%%%%%%%%%%%%%
%%%%%%%%%%%%%%%%%%%%%%%%
\begin{figure}[p]
\hspace{\stretch{1}}\begin{subfigure}{0.3\textwidth}
\includegraphics[width=120pt]{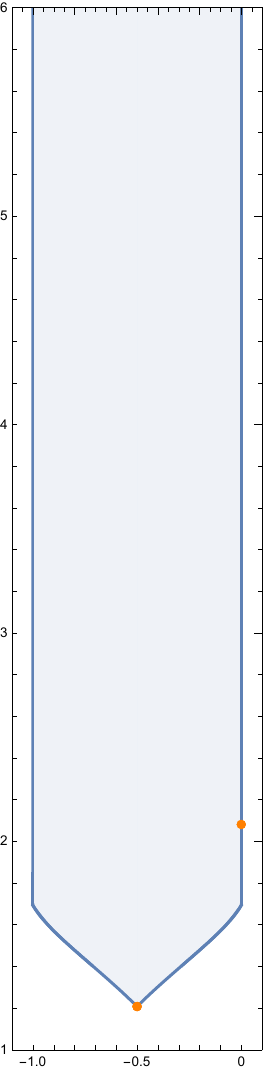}
\caption{$N_{\rm flux}=1$.}
\end{subfigure}
\hspace{\stretch{1}}
\begin{subfigure}{0.3\textwidth}
\includegraphics[width=120pt]{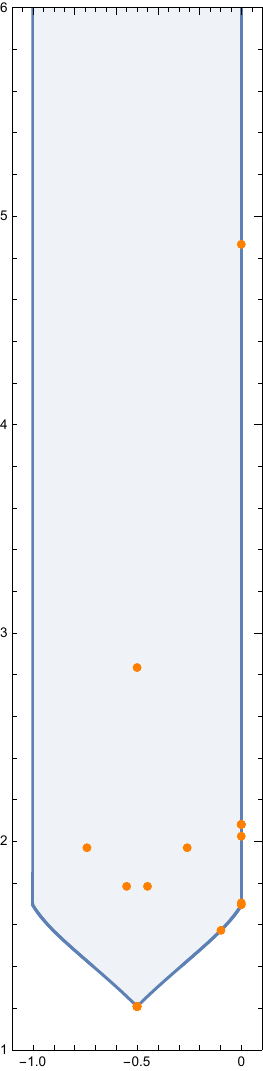}
\caption{$N_{\rm flux}=2$.}
\end{subfigure}
\hspace{\stretch{1}}
\begin{subfigure}{0.3\textwidth}
\includegraphics[width=120pt]{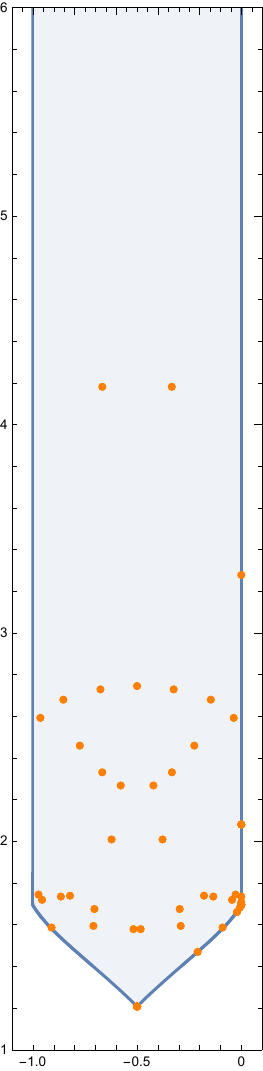}
\caption{$N_{\rm flux}=3$.}
\end{subfigure}
\hspace{\stretch{1}}
\end{figure}
%%%%%%%%%%%%%%%%%%%%%%%%
%%%%%%%%%%%%%%%%%%%%%%%%
\begin{figure}[p]
\hspace{\stretch{1}}\begin{subfigure}{0.3\textwidth}
\includegraphics[width=120pt]{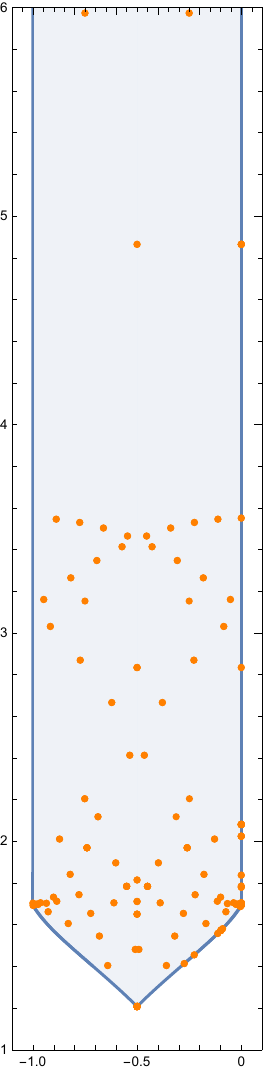}
\caption{$N_{\rm flux}=4$.}
\end{subfigure}
\hspace{\stretch{1}}
\begin{subfigure}{0.3\textwidth}
\includegraphics[width=120pt]{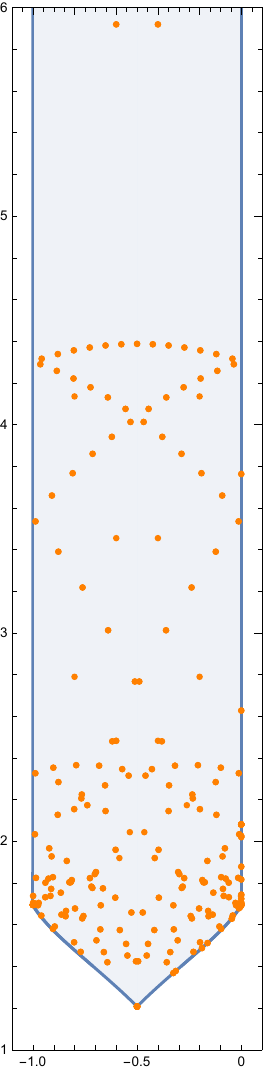}
\caption{$N_{\rm flux}=5$.}
\end{subfigure}
\hspace{\stretch{1}}
\begin{subfigure}{0.3\textwidth}
\includegraphics[width=120pt]{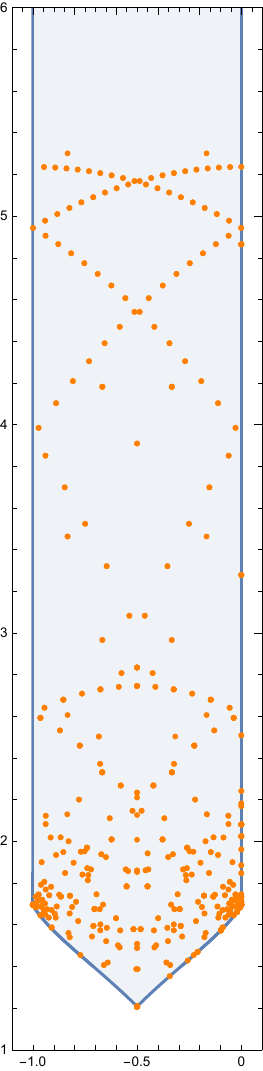}
\caption{$N_{\rm flux}=6$.}
\end{subfigure}
\hspace{\stretch{1}}
\end{figure}
%%%%%%%%%%%%%%%%%%%%%%%%
%%%%%%%%%%%%%%%%%%%%%%%%
\begin{figure}[p]\ContinuedFloat
\hspace{\stretch{1}}\begin{subfigure}{0.3\textwidth}
\includegraphics[width=120pt]{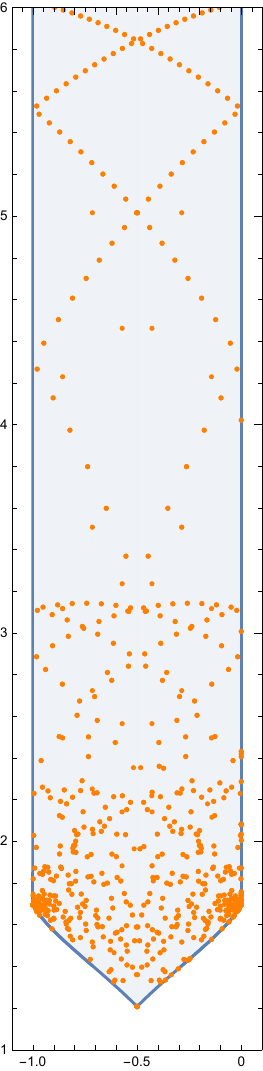}
\caption{$N_{\rm flux}=7$.}
\end{subfigure}
\hspace{\stretch{1}}
\begin{subfigure}{0.3\textwidth}
\includegraphics[width=120pt]{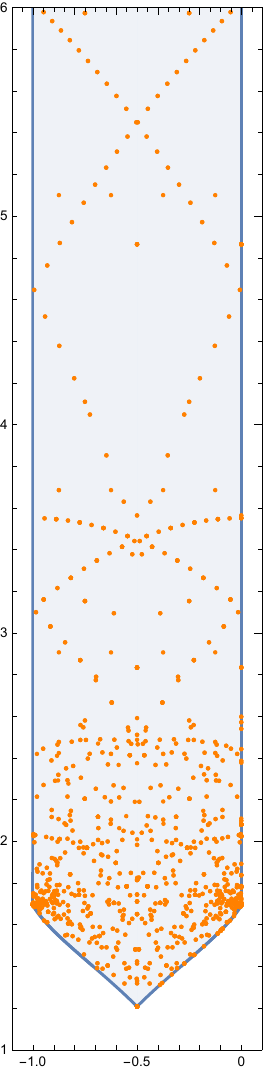}
\caption{$N_{\rm flux}=8$.}
\end{subfigure}
\hspace{\stretch{1}}
\begin{subfigure}{0.3\textwidth}
\includegraphics[width=120pt]{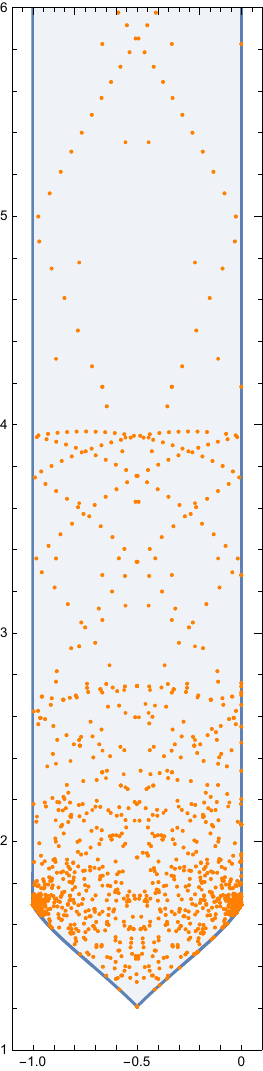}
\caption{$N_{\rm flux}=9$.}
\end{subfigure}
\hspace{\stretch{1}}
\end{figure}
%%%%%%%%%%%%%%%%%%%%%%%%
%%%%%%%%%%%%%%%%%%%%%%%%
\begin{figure}[p]\ContinuedFloat
\vspace{40pt}
\hspace{\stretch{1}}\begin{subfigure}{0.3\textwidth}
\includegraphics[width=120pt]{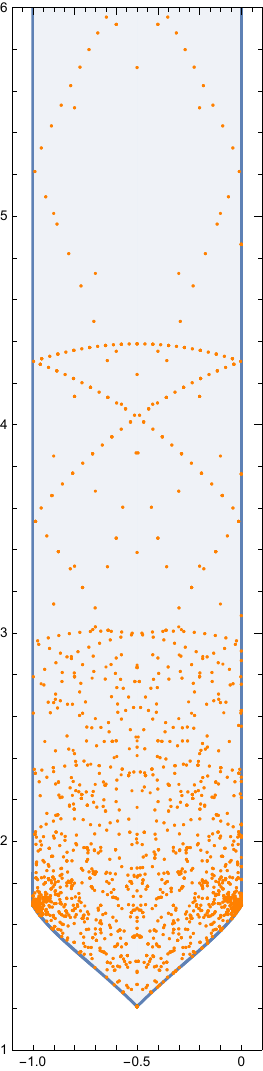}
\caption{$N_{\rm flux}=10$.}
\end{subfigure}
\hspace{\stretch{1}}
\caption{Distribution of stabilized complex-structure moduli on the K\"ahler side
for different values of the flux number $N_{\rm flux}$. \label{fig_minima}}
\end{figure}
%%%%%%%%%%%%%%%%%%%%%%%%
%%%%%%%%%%%%%%%%%%%%%%%%

%%%%%%%%%%%%%%%%%%%%%%%%
%%%%%%%%%%%%%%%%%%%%%%%%
%%%%%%%%%%%%%%%%%%%%%%%%
%%%%%%%%%%%%%%%%%%%%%%%%

%%%%%%%%%%%%%%%%%%%%%%%%%%%%%%%%%%%%%%%%%%%%%%%
%%%%%%%%%%%%%%%%%%%%%%%%%%%%%%%%%%%%%%%%%%%%%%%
%%%%%%%%%%%%%%%%%%%%%%%%%%%%%%%%%%%%%%%%%%%%%%%
%%%%%%%%%%%%%%%%%%%%%%%%%%%%%%%%%%%%%%%%%%%%%%%
%%%%%%%%%%%%%%%%%%%%%%%%%%%%%%%%%%%%%%%%%%%%%%%

\subsection{Number of vacua}

In this subsection we analyze and discuss how the
number of flux vacua $\mathcal N$ for the mirror octic depends on
the flux number $N_{\rm flux}$. 

%%%%%%%%%%%%%%%%%%%%%%%%%%%%%%%%%%%%%%%%%%%%%%%
%%%%%%%%%%%%%%%%%%%%%%%%%%%%%%%%%%%%%%%%%%%%%%%

\subsubsection*{Results}

\begin{figure}[p]
\centering
\includegraphics[width=280pt]{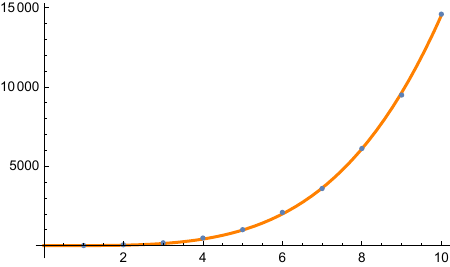}%
\begin{picture}(0,0)
\put(-23,-8){\footnotesize $N_{\rm max}$}
\put(-301,168){\footnotesize \# of vacua}
\put(-92,130){\footnotesize $  1.88 \cdot \bigl(N_{\rm max}\bigr)^{3.89}$}
\end{picture}
\vspace*{10pt}
\caption{Number of vacua that satisfy $N_{\rm flux}\leq N_{\rm max}$.
\label{fig_number_vacua}}
\end{figure}

\begin{figure}[p]
\centering
\includegraphics[width=280pt]{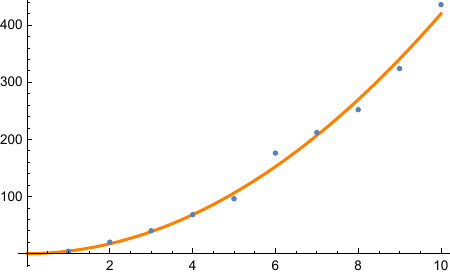}%
\begin{picture}(0,0)
\put(-23,-8){\footnotesize $N_{\rm max}$}
\put(-312,174){\footnotesize \# of vacua}
\put(-108,130){\footnotesize $  4.32 \cdot \bigl(N_{\rm max}\bigr)^{1.99}$}
\end{picture}
\vspace*{10pt}
\caption{Number of vacua with $W_0=0$ that satisfy $N_{\rm flux}\leq N_{\rm max}$.
\label{fig_number_vacua_w0}}
\end{figure}

We first note that the scalar potential \eqref{spot} and the minimum conditions \eqref{min_99}
are invariant under the following change of sign of the flux vectors
\eq{
  \label{rel_1020}
  (\op\mathsf H,\mathsf F\op) \to (-\mathsf H,-\mathsf F\op)\,.
}
We therefore consider two flux choices that are related by 
\eqref{rel_1020} as equivalent. 
The numbers of vacua for a fixed $N_{\rm flux}$ 
have been summarized in table~\ref{tab_number_vacua_1}
on page~\pageref{page_table_1},
and the  cumulative 
number of vacua $\mathcal N(N_{\rm flux}\leq N_{\rm max})$ for which the flux number satisfies 
$N_{\rm flux}\leq N_{\rm max}$ are shown in table~\ref{tab_number_vacua_2}.
The cumulative data is shown in figure~\ref{fig_number_vacua} and can be fitted
as follows
\eq{
  \label{fit_numbers}
   \mathcal N \bigl(N_{\rm flux}\leq N_{\rm max}\bigr)_{\hphantom{W_0=0}} =   1.88 \cdot \bigl(N_{\rm max}\bigr)^{3.89}\,,
}
and the cumulative data for vacua with $W_0=0$ is shown in figure~\ref{fig_number_vacua_w0} and 
is  approximated by the following function
\eq{
  \label{fit_numbers_w0}
  \mathcal N \bigl(N_{\rm flux}\leq N_{\rm max}\bigr)_{W_0=0} =   4.32 \cdot \bigl(N_{\rm max}\bigr)^{1.99}\,.
}
Note that all vacua with $W_0=0$ are  located at the 
Landau-Ginzburg point $z=\infty$.

%%%%%%%%%%%%%%%%%%%%%%%%
%%%%%%%%%%%%%%%%%%%%%%%%
%%%%%%%%%%%%%%%%%%%%%%%%
%%%%%%%%%%%%%%%%%%%%%%%%

%%%%%%%%%%%%%%%%%%%%%%%%
%%%%%%%%%%%%%%%%%%%%%%%%
%%%%%%%%%%%%%%%%%%%%%%%%
%%%%%%%%%%%%%%%%%%%%%%%%

%%%%%%%%%%%%%%%%%%%%%%%%%%%%%%%%%%%%%%%%%%%%%%%
%%%%%%%%%%%%%%%%%%%%%%%%%%%%%%%%%%%%%%%%%%%%%%%

\subsubsection*{Comparison with estimates by Denef/Douglas}

We now want to compare our results for the number of vacua shown 
in \eqref{fit_numbers} with the estimates 
in the literature \cite{
Ashok:2003gk,
Denef:2004ze
}.
In particular, in \cite{Denef:2004ze}  a vacuum density $d\mu$ 
and a vacuum index density $d\mu_{\rm ind}$ are defined which 
can be integrated to give an estimate on the  number of vacua.
For type IIB orientifolds with  $h^{2,1}_-=1$  we recall from 
equations (3.19) and (3.20) of  \cite{Denef:2004ze}  the expressions
\eq{
  \rho = \frac{1}{\pi^2} \left( 2- \lvert \mathcal F \rvert^2+
  \frac{2\op \lvert \mathcal F \rvert^3}{\sqrt{+ \lvert \mathcal F \rvert^2}}
  \right),
  \hspace{40pt}
  \rho_{\rm ind} = \frac{1}{\pi^2} \left( 2- \lvert \mathcal F \rvert^2
  \right),
}
where $\mathcal F$ is defined as
\eq{
  \mathcal F = \mathcal G_{z\ov z}^{-3/2}\op e^{\mathcal K_{\rm cs}}
  \left( +i\, \Pi^T \eta\, \partial_z^3 \Pi \right) .
}
The corresponding vacuum densities 
are given by  $d\mu = d^2\tau\op  d^2 z\, g_{\tau\ov\tau} \op g_{z\ov z} \op \rho$
and by $d\mu_{\rm ind} = d^2\tau\op  d^2 z\, g_{\tau\ov\tau} \op  \op g_{z\ov z} \op \rho_{\rm ind}$
and are understood to be integrated over the combined moduli space $\mathsf M$ of the axio-dilaton $\tau$ 
and complex-structure modulus $z$. 
Partitioning the latter according to the covering shown in \eqref{regions_finite}
and \eqref{regions_sing}, we can perform the integral numerically. 
We find for the mirror octic
\eq{
  \label{int_1031}
  &\int_{\mathsf M} d\mu_{\hphantom{\rm ind}} = \int_{\mathsf M_{\tau}} d^2\tau\op g_{\tau\ov \tau}
  \int_{\mathsf M_{\rm cs}} d^2z\op g_{z\ov z} \, \rho_{\hphantom{\rm ind}} = \hphantom{+}4.22\cdot 10^{-2} \,,
  \\[8pt]
  &\int_{\mathsf M} d\mu_{\rm ind} = \int_{\mathsf M_{\tau}} d^2\tau\op g_{\tau\ov \tau}
  \int_{\mathsf M_{\rm cs}} d^2z\op g_{z\ov z} \, \rho_{\rm ind} = -2.01\cdot 10^{-2} \,,
}
where the integral over the axio-dilaton moduli space contributes a factor of $\pi/12$.
These results are  similar to the numbers obtained for the mirror quintic 
(see for instance equation (3.33) and (3.34) of \cite{Denef:2004ze}).\footnote{
By the same reasoning as for the mirror quintic, the 
result for the index density suggests that 
the integral of the Euler class for the mirror octic is close to $\chi(\mathcal X) = -1/4$.
}
The estimated number of vacua for our setting can then be determined as follows
\cite{Denef:2004ze}
\eq{
  \mathcal N_{\rm est} \bigl(N_{\rm flux}\leq N_{\rm max}\bigr) = \frac{\bigl( 2\pi N_{\rm max}\bigr)^{4}}{4!} \int_{\mathsf M} d\mu
}
and similarly (with absolute values) for the index density. Using  the numbers shown in \eqref{int_1031}, we find
\begin{subequations}
\label{est_number}
\begin{align}
\label{est_number_a}
    &\mathcal N_{{\rm est}\hphantom{\lvert{\rm ind}}} \bigl(N_{\rm flux}\leq N_{\rm max}\bigr)  
    =     2.74\cdot \bigl(N_{\rm max}\bigr)^{4} \,,
    \\
\label{est_number_b}    
    &\mathcal N_{{\rm est}\lvert{\rm ind}} \bigl(N_{\rm flux}\leq N_{\rm max}\bigr)  
    =     1.31\cdot \bigl(N_{\rm max}\bigr)^{4} \,.
\end{align}
\end{subequations}
We compare these expression to our results:
\begin{itemize}

\item We first note that the scaling exponent $3.89$ in our result \eqref{fit_numbers} matches 
approximately  the estimated value of four shown in \eqref{est_number}.

\item For the prefactor in   \eqref{fit_numbers} we  recall that 
we consider two flux-confi\-gu\-rations to be equivalent if they satisfy \eqref{rel_1020}.
In order to compare our fit to 
\eqref{est_number}, we need to include an additional factor of two and 
hence obtain for the prefactor $3.76$.
We therefore have
\eq{
  \begin{array}{l@{\hspace{40pt}}r}
  \mbox{prefactor of fit to data in \eqref{fit_numbers}:} & 2\cdot 1.88 = 3.76\,,
  \\[6pt]
  \mbox{prefactor of estimate in \eqref{est_number_a}:} &  2.74\,.
  \end{array}
}
The small discrepancy between our result and 
the estimate by \cite{Denef:2004ze} could be explained by 
the comparable numbers $h^{2,1}_-$ and $N_{\rm flux}$ we are considering, for which  
the continuous-flux approximation 
of  \cite{Denef:2004ze} might not be a good approximation.

\item We also note that \eqref{est_number_b} is only a lower bound on 
the number of vacua and therefore it is compatible with our data.

\end{itemize}

%%%%%%%%%%%%%%%%%%%%%%%%%%%%%%%%%%%%%%%%%%%%%%%
%%%%%%%%%%%%%%%%%%%%%%%%%%%%%%%%%%%%%%%%%%%%%%%

\subsubsection*{Comparison with results on $W_0=0$ vacua}

The vacua for which the superpotential vanishes at the minimum are all located at the 
Landau-Ginzburg point.
The dependence of number of vacua with $W_0$ on the flux number is summarized in tables~\ref{tab_number_vacua_1}
and \ref{tab_number_vacua_2},  and 
the cumulative number of vacua for $W_0=0$ has been shown 
in figure~\ref{fig_number_vacua_w0} and has been fitted 
in equation \eqref{fit_numbers_w0}.
Let us compare these results to the literature:
 the authors of 
\cite{DeWolfe:2004ns} 
found that the  number of vacua at the Landau-Ginzburg point of the mirror-octic 
 $\mathcal N_{\rm LG} (N_{\rm flux}\leq N_{\rm max})$ scales as
 $(N_{\max})^2$, which agrees with the scaling we find in  \eqref{fit_numbers_w0}.
In \cite{DeWolfe:2004ns} also vacua 
with $W_0=0$ were  investigated,  but the authors did not find such vacua at the LG point. 
We, however, do find such solutions in our search.

%%%%%%%%%%%%%%%%%%%%%%%%%%%%%%%%%%%%%%%%%%%%%%%
%%%%%%%%%%%%%%%%%%%%%%%%%%%%%%%%%%%%%%%%%%%%%%%
%%%%%%%%%%%%%%%%%%%%%%%%%%%%%%%%%%%%%%%%%%%%%%%
%%%%%%%%%%%%%%%%%%%%%%%%%%%%%%%%%%%%%%%%%%%%%%%
%%%%%%%%%%%%%%%%%%%%%%%%%%%%%%%%%%%%%%%%%%%%%%%

\subsection{Distribution of $W_0$}

Next, we discuss the distribution of the value of the superpotential 
at the minimum.
As explained around equation \eqref{superpot_200},
we consider the following two expressions
\eq{
\label{w0_100}
  e^{\mathcal K_{\rm cs}}\lvert W_0\rvert^2 \,,
  \hspace{80pt}
  e^{\mathcal K_{\tau}+\mathcal K_{\rm cs}}\lvert W_0\rvert^2 \,.
}

%%%%%%%%%%%%%%%%%%%%%%%%%%%%%%%%%%%%%%%%%%%%%%%
%%%%%%%%%%%%%%%%%%%%%%%%%%%%%%%%%%%%%%%%%%%%%%%

\subsubsection*{Discussion for $e^{\mathcal K_{\rm cs}}\lvert W_0\rvert^2$}

\begin{figure}[p]
\centering
\includegraphics[width=280pt]{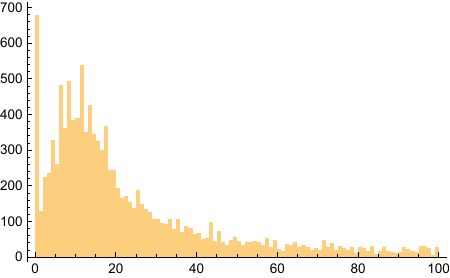}%
\begin{picture}(0,0)
\put(-42,-11){\footnotesize $e^{\mathcal K_{\rm cs}}\lvert W_0\rvert^2$}
\put(-287,174){\footnotesize count}
\end{picture}
\vspace*{10pt}
\caption{Histogram of $e^{\mathcal K_{\rm cs}}\lvert W_0\rvert^2$ 
(including $W_0=0$) up to the value $100$
for $N_{\rm flux}\leq 10$.
Note that the maximal value 
is $e^{\mathcal K_{\rm cs}}\lvert W_0\rvert^2\rvert_{\rm max}=2.41 \cdot 10^8$
and that the left-most bin contains the vacua with $W_0=0$.
The bin-size $\mathsf b$ is chosen as $\mathsf b=1$.
\label{fig_w0_1}}
\end{figure}

The first expression in \eqref{w0_100} is relevant for K\"ahler-moduli stabilization in the 
KKLT \cite{Kachru:2003aw} and Large-Volume Scenarios (LVS) \cite{Balasubramanian:2005zx}. 
For KKLT the value of  $e^{\mathcal K_{\rm cs}/2}\lvert W_0\rvert$ should be much less than one, 
while for LVS it can be of order one.
The distribution of $e^{\mathcal K_{\rm cs}}\lvert W_0\rvert^2$ for the mirror-octic 
is shown 
in figure~\ref{fig_w0_1}, and we make the following observations:
\begin{itemize}

\item We were not able to find an analytic expression 
for the  distribution of $e^{\mathcal K_{\rm cs}}\lvert W_0\rvert^2$.
Though, as can be inferred from 
figure~\ref{fig_w0_1},
it is not a normal distribution.

\item If we exclude the vacua with $W_0=0$, 
we obtain the minimal values 
of $e^{\mathcal K_{\rm cs}}\lvert W_0\rvert^2$ shown in table~\ref{table_w0_min_max}.
These values are not exponentially-small, and hence stabilization of K\"ahler moduli 
via KKLT will be difficult for this model.

\item 
For completeness we also show in table~\ref{table_w0_min_max} 
the maximal values of $e^{\mathcal K_{\rm cs}}\lvert W_0\rvert^2$ for each flux number.
Furthermore, vacua with $e^{\mathcal K_{\rm cs}/2}\lvert W_0\rvert$ of 
order one and weak string-coupling $s\geq 5$ can be found,
which  allows for 
K\"ahler-moduli stabilization via LVS.

\end{itemize}
%%%%%%%%%%%%%%%%%%%%%%%%
%%%%%%%%%%%%%%%%%%%%%%%%
%%%%%%%%%%%%%%%%%%%%%%%%
%%%%%%%%%%%%%%%%%%%%%%%%

%%%%%%%%%%%%%%%%%%%%%%%%
%%%%%%%%%%%%%%%%%%%%%%%%
\begin{figure}[p]
\centering
\includegraphics[width=280pt]{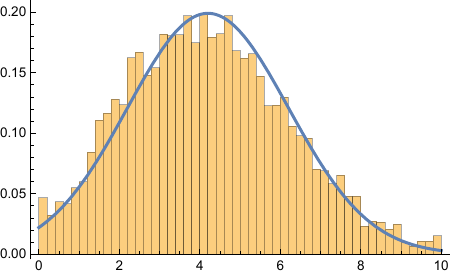}%
\begin{picture}(0,0)
\put(-57,-11){\footnotesize $e^{\mathcal K_{\tau}+\mathcal K_{\rm cs}}\lvert W_0\rvert^2$}
\put(-309,176){\footnotesize probability}
\end{picture}
\vspace*{10pt}
\caption{Probability distribution of $e^{\mathcal K_{\tau}+\mathcal K_{\rm cs}}\lvert W_0\rvert^2$ for $N_{\rm flux}\leq 10$
excluding vacua with $W_0=0$.
The bin-size $\mathsf b$ is chosen as $\mathsf b=\frac15$.
\label{fig_w0_2}}
\end{figure}
%%%%%%%%%%%%%%%%%%%%%%%%
%%%%%%%%%%%%%%%%%%%%%%%%

%%%%%%%%%%%%%%%%%%%%%%%%
%%%%%%%%%%%%%%%%%%%%%%%%
\begin{table}[p]
\hspace{\stretch{1}}
\begin{subtable}{.45\textwidth}
\centering
\eq{
  \nonumber
  \renewcommand{\arraystretch}{1.2}
  \arraycolsep10pt
  \begin{array}{@{}|| c || c || c ||@{}}
  \hline\hline
  \multirow{2}{*}{$N_{\rm flux}$} & \multicolumn{2}{c ||}{\mbox{\# of vacua}} \\
  \cline{2-3} 
  & \mbox{all $W_0$} & W_0=0
  \\
  \hline \hline
  1 & 6 & 4\\
  2 & 46 & 16\\
  3 & 123 & 20\\
  4 & 302 & 28\\
  5 & 528 &28 \\
  6 & 1089 & 80\\
  7 & 1503 & 36\\
  8 & 2522 & 40\\
  9 & 3368 & 72\\
  10 & 5097 &112 \\
  \hline\hline
  \end{array}
}
\caption{Number of vacua for each $N_{\rm flux}$.\label{tab_number_vacua_1}}
\end{subtable}
\hspace{\stretch{1}}
\begin{subtable}{.45\textwidth}
\centering
\eq{
  \nonumber
  \renewcommand{\arraystretch}{1.2}
  \arraycolsep10pt
 \begin{array}{@{}|| c || c || c ||@{}}
  \hline\hline
  \multirow{2}{*}{$N_{\rm max}$} & \multicolumn{2}{c ||}{\mbox{\# of vacua}} \\
  \cline{2-3} 
  & \mbox{all $W_0$} & W_0=0
  \\
  \hline \hline
  1 & 6 & 4\\
  2 & 52 & 20 \\
  3 & 175 & 40\\
  4 & 477 & 68\\
  5 & 1005 & 96 \\
  6 & 2094 & 176\\
  7 & 3597 & 212\\
  8 & 6119 & 252\\
  9 & 9487 & 324\\
  10 & 14584 & 436\\
  \hline\hline
  \end{array}
}
\caption{Number of vacua for $N_{\rm flux}\leq N_{\rm max}$.\label{tab_number_vacua_2}}
\end{subtable}
\hspace{\stretch{1}}
\caption{
Individual and cumulative number of flux vacua for the mirror octic.\label{page_table_1}
}
\end{table}
%%%%%%%%%%%%%%%%%%%%%%%%
%%%%%%%%%%%%%%%%%%%%%%%%

%%%%%%%%%%%%%%%%%%%%%%%%
%%%%%%%%%%%%%%%%%%%%%%%%
\begin{table}[p]
\eq{
\nonumber
  \renewcommand{\arraystretch}{1.2}
  \arraycolsep10pt
  \begin{array}{|| c||c || c ||}
  \hline\hline
  N_{\rm flux} & \mbox{min}_{W_0\neq 0}\bigl( e^{\mathcal K_{\rm cs}}\lvert W_0\rvert^2 \bigr)
  & \mbox{max}\bigl( e^{\mathcal K_{\rm cs}}\lvert W_0\rvert^2 \bigr)
  \\
  \hline \hline
  1 & 1.0889 & 3.96\cdot 10^0\\
  2 & 0.7241 & 4.84 \cdot 10^3\\
  3 & 0.7167 & 3.68 \cdot 10^4\\
  4 & 0.0247 & 1.24\cdot 10^{6}\\
  5 & 0.0029 & 4.73 \cdot 10^{5}\\
  6 & 0.0717 & 1.47\cdot 10^{5}\\
  7 & 0.0839 & 6.55 \cdot 10^{5}\\
  8 & 0.0247 & 4.96\cdot 10^6\\
  9 & 0.0097 & 2.41 \cdot 10^8\\
10 & 0.0029& 5.51\cdot 10^6 \\
  \hline\hline
  \end{array}
}
\caption{Minimal  and maximal values of $ e^{\mathcal K_{\rm cs}}\lvert W_0\rvert^2 $
for each flux number $N_{\rm flux}$.
We exclude vacua with $W_0=0$.
\label{table_w0_min_max}
}
\end{table}
%%%%%%%%%%%%%%%%%%%%%%%%
%%%%%%%%%%%%%%%%%%%%%%%%

%%%%%%%%%%%%%%%%%%%%%%%%%%%%%%%%%%%%%%%%%%%%%%%
%%%%%%%%%%%%%%%%%%%%%%%%%%%%%%%%%%%%%%%%%%%%%%%

\subsubsection*{Discussion for $e^{\mathcal K_{\tau}+\mathcal K_{\rm cs}}\lvert W_0\rvert^2$}

The second expression in \eqref{w0_100} corresponds to the gravitino-mass squared 
--- up to an overall volume factor $1/\mathcal V^2$
that is only fixed after the K\"ahler moduli are stabilized. Here we find
that 
\eq{
e^{\mathcal K_{\tau}+\mathcal K_{\rm cs}}\lvert W_0\rvert^2 \leq N_{\rm flux}\,,
}
which can be proven on general grounds using the second relation in \eqref{min_102}.
If we exclude the vacua with $W_0=0$, the distribution of 
$e^{\mathcal K_{\tau}+\mathcal K_{\rm cs}}\lvert W_0\rvert^2$ for our dataset 
can be fitted by a normal 
distribution with mean and standard deviation
\eq{
  \mu=0.427\op N_{\rm max}\,,
  \hspace{60pt}
  \sigma = 0.204\op N_{\rm max}\,,
}
respectively. This is illustrated in figure~\ref{fig_w0_2} for $N_{\rm max}=10$.
We also note that the authors of \cite{Ebelt:2023clh} found 
for the complex quantity $e^{(\mathcal K_{\tau}+\mathcal K_{\rm cs})/2} W_0$ 
a zero-mean, isotropic, bivariante, normal distribution
in their examples.
This implies a Rayleigh distribution for $e^{(\mathcal K_{\tau}+\mathcal K_{\rm cs})/2} \lvert W_0\rvert$,
which however does not fit our data. We do not have an explanation for  
this difference.\footnote{We thank S.~Krippendorf and A.~Schachner for 
discussions on this matter.} \label{page_ebelt}

%%%%%%%%%%%%%%%%%%%%%%%%%%%%%%%%%%%%%%%%%%%%%%%
%%%%%%%%%%%%%%%%%%%%%%%%%%%%%%%%%%%%%%%%%%%%%%%
%%%%%%%%%%%%%%%%%%%%%%%%%%%%%%%%%%%%%%%%%%%%%%%
%%%%%%%%%%%%%%%%%%%%%%%%%%%%%%%%%%%%%%%%%%%%%%%
%%%%%%%%%%%%%%%%%%%%%%%%%%%%%%%%%%%%%%%%%%%%%%%

\subsection{Sampling vs. complete scan}

We recall that a fixed flux number $N_{\rm flux}$ can be realized by infinitely-many
flux vectors $\mathsf H$ and $\mathsf F$.
Without imposing the bounds derived in section~\ref{sec_bounds} --- which, 
to our knowledge,  have not appeared in this context in the literature before ---
one often needs to sample flux vectors randomly 
from a box defined as
\eq{
  h^I,\op h_I,\op f^I,\op f_I \in [-L_{\rm box},+L_{\rm box}]\,.
}
However, as developed in \cite{Tsagkaris:2022apo}
and later advanced in \cite{Dubey:2023dvu},
one can restrict the sampling of the fluxes for instance in the LCS region.
In this subsection we want to ask the question
what fraction of all flux vacua is contained in 
a box specified by $L_{\rm box}$? 
To address this point, we first define the ratio of the number of vacua in a box of size $L_{\rm box}$ to the
number of all flux vacua with $N_{\rm flux}\leq N_{\rm max}$ as
\eq{
  \label{fraction_1}
  r(L_{\rm box},N_{\rm max})= \frac{\mbox{\# of vacua contained in a box of size $L_{\rm box}$}}
  {  \mbox{\# of all flux vacua}}\biggr\rvert_{N_{\rm flux}\leq N_{\rm max}}\,.
}
We determined $  r(L_{\rm box},N_{\rm max})$ for our setting and have summarized this data  in table~\ref{table_fraction}.
%%%%%%%%%%%%%%%%%%%%%%%%
%%%%%%%%%%%%%%%%%%%%%%%%
\begin{table}[p]
\eq{
\nonumber
\renewcommand{\arraystretch}{1.2}
\arraycolsep6.25pt
\begin{array}{|| c || ccccccccc ||}
\hline\hline
$\diagbox[width=1.6cm,height=1.6cm]
{\footnotesize{$N_{\rm max}$}}
{\footnotesize{$L_{\rm box}$}}
$ 
& 1 & 2 & 3 & 4 & 5 &10 & 15 & 25 & 35
\\
\hline\hline
1 & 0.667 & 1 & 1 & 1 & 1 & 1 & 1 & 1 & 1
\\
2 & 0.173 & 0.808 & 0.962 & 1 & 1 & 1 & 1& 1& 1
\\
3 & 0.051 & 0.475 & 0.814 & 0.898 & 0.960 & 1 & 1 & 1 & 1
\\
4 & 0.019 & 0.344 & 0.629 & 0.838 & 0.902 & 0.996 & 1 & 1 & 1 
\\
5 & 0.009 & 0.211 & 0.483&0.691&0.826&0.982&1&1&1
\\
6 &0.004&0.137&0.391&0.604&0.744&0.968&0.998&1&1
\\
7 & 0.002&0.087&0.305&0.503&0.659&0.942&0.991&1&1
\\
8 & 0.001&0.060&0.239&0.443&0.588&0.917&0.983&1&1
\\
9 &0.001&0.040&0.199&0.378&0.526&0.886&0.970&0.999&1
\\
10 &0.001&0.026&0.156&0.328&0.474&0.860&0.958&0.997&\hspace*{8pt}1\hspace*{8pt}
\\
\hline\hline
\end{array}
}
\vspace*{-10pt}
\caption{The ratio $r(L_{\rm box},N_{\rm max})$ defined in equation \eqref{fraction_1}
for different values of $L_{\rm box}$ and  $N_{\rm max}$.
\label{table_fraction}
}
\end{table}
%%%%%%%%%%%%%%%%%%%%%%%%
%%%%%%%%%%%%%%%%%%%%%%%%
Let us discuss one  example from that table:
\begin{itemize}

\item From table~\ref{table_fraction} we find  that
$82.6\%$ of the vacua with flux numbers $N_{\rm flux}\leq 5$ 
are contained within a box of size $L_{\rm box}=5$.

\item For our model a box with $L_{\rm box}=5$ contains $\frac12 \cdot 8^{2\cdot 5 + 1} \simeq 4.3\cdot 10^9$ 
flux combinations while there are only $837$  flux vacua with $N_{\rm flux}\leq 5$ within this box.
The ratio of flux vacua to all flux choices for $L_{\rm box}=5$ is therefore $1.9\cdot 10^{-7}$.

\end{itemize}
From the ratios summarized in table~\ref{table_fraction} we can conclude that  
1) only a  fraction of flux vacua can in principle be found when restricting the flux quanta to lie 
within a moderately-sized box, and that 2) finding these flux vacua via a random search will 
be difficult.

%%%%%%%%%%%%%%%%%%%%%%%%%%%%%%%%%%%%%%%%%%%%%%%
%%%%%%%%%%%%%%%%%%%%%%%%%%%%%%%%%%%%%%%%%%%%%%%
%%%%%%%%%%%%%%%%%%%%%%%%%%%%%%%%%%%%%%%%%%%%%%%
%%%%%%%%%%%%%%%%%%%%%%%%%%%%%%%%%%%%%%%%%%%%%%%
%%%%%%%%%%%%%%%%%%%%%%%%%%%%%%%%%%%%%%%%%%%%%%%

\subsection{Numerics}

In this section we comment on 
 the estimated CPU hours, on the structure of the dataset that has been submitted 
 with this paper to the \texttt{arXiv}, and on numerical errors.
\begin{itemize}

\item Most of the computations to determine the set of flux choices $(\mathsf H,\mathsf F)$ 
and to solve the minimum conditions \eqref{rel_401} have been done on a computer 
cluster at Utrecht University. We estimate the total CPU time for this project as $3300\op\mbox{hrs}$,
and the CPU time for obtaining our data set as
\eq{
  \mbox{CPU time} \simeq 2000\op\mbox{hrs} \simeq 83\op\mbox{days} \,.
}

\item The dataset of all flux vacua for the mirror-octic with $N_{\rm flux}\leq 10$ can be
found on this paper's \texttt{arXiv} page. To obtain this dataset
choose \texttt{Other Formats}, download the source of the submission, add the extension \texttt{.zip} to the file, 
unzip the file, and open the folder \texttt{\textbackslash anc}.
The data is contained in the file 
\texttt{data\_octic.csv} and has the structure shown in table~\ref{table_dataset}.

\item In this paper we have constructed all flux vacua for the mirror octic --- up to numerical errors. 
To explain this point, let us recall that our scan for  vacua consists of two steps, namely: 1) constructing a finite set of flux 
configurations and 2) solving the F-term equations for each flux choice. 

In regard to the first step, we note that the bounds on the fluxes derived in section~\ref{sec_bounds} ensure that 
in the finite regions of moduli space we capture all relevant flux choices. For the boundary regions with 
$v\geq 3$, in particular for 
the LCS region with $\mathsf h_1=0$ and for the conifold region, we believe that the approximations 
we made for the period vector are sufficient to capture all relevant flux choices in these regimes.
Higher-order exponential corrections typically lead to contributions below the numerical precision
and are therefore not important for our scan.

In regard to the second step, we mention  that a minimum of the scalar potential may not be found by 
\texttt{Mathematica}'s search algorithm. We limited this possibility by having 30 randomly-chosen starting points for 
the minimization algorithm --- increasing this number did not lead to additional solutions. We have also performed various cross checks for instance on the overlap of two 
patches in complex-structure moduli space (c.f.~figure~\ref{FundamentalDomainCover}).

\end{itemize}

%%%%%%%%%%%%%%%%%%%%%%%%
%%%%%%%%%%%%%%%%%%%%%%%%
\begin{table}[p]
\centering
\renewcommand{\arraystretch}{1.2}
\tabcolsep8pt
\begin{tabular}{|| c || c || l ||}
\hline\hline
column & data & comments \\
\hline\hline
1 - 4 & $\mathsf H$ & flux vector defined in \eqref{rel_270}
\\
5 - 8 & $\mathsf F$ & flux vector defined in \eqref{rel_270}
\\
9 & $N_{\rm flux}$ & flux number $N_{\rm flux} = \mathsf F^T \eta \op \mathsf H$
\\
10 & $\mbox{Re}\op(t)$ & real part of the K\"ahler coordinate \eqref{coord_kaehler}
\\
11 & $\mbox{Im}\op(t)$ & imaginary part of the K\"ahler coordinate \eqref{coord_kaehler}
\\
12 & $z_{\rm ep}$ & expansion point of the periods, c.f.~\eqref{regions_finite}
\\
13 & $u$ & local variable for expansion of periods, c.f.~\eqref{coord_exp}
\\
14 & $v$ & local variable for expansion of periods, c.f.~\eqref{coord_exp}
\\
15 & $c$ & real part of axio-dilaton $\tau$, c.f.~\eqref{coord_tau}
\\
16 & $s$ & imaginary part of axio-dilaton $\tau$, c.f.~\eqref{coord_tau}
\\
17 & $ \hspace{18pt}e^{\mathcal K_{\rm cs}}\lvert W_0\rvert^2 $ & value of superpotential at minimum, c.f.~\eqref{superpot_200}
\\
18 & $ e^{\mathcal K_{\tau}+\mathcal K_{\rm cs}}\lvert W_0\rvert^2 $ & value of superpotential at minimum, c.f.~\eqref{superpot_200}
\\
\hline\hline
\end{tabular}
\caption{Structure of the dataset.\label{table_dataset}}
\end{table}
%%%%%%%%%%%%%%%%%%%%%%%%
%%%%%%%%%%%%%%%%%%%%%%%%

%%%%%%%%%%%%%%%%%%%%%%%%%%%%%%%%%%%%%%%%%%%%%%%
%%%%%%%%%%%%%%%%%%%%%%%%%%%%%%%%%%%%%%%%%%%%%%%
%%%%%%%%%%%%%%%%%%%%%%%%%%%%%%%%%%%%%%%%%%%%%%%
%%%%%%%%%%%%%%%%%%%%%%%%%%%%%%%%%%%%%%%%%%%%%%%
%%%%%%%%%%%%%%%%%%%%%%%%%%%%%%%%%%%%%%%%%%%%%%%
%%%%%%%%%%%%%%%%%%%%%%%%%%%%%%%%%%%%%%%%%%%%%%%
%%%%%%%%%%%%%%%%%%%%%%%%%%%%%%%%%%%%%%%%%%%%%%%
%%%%%%%%%%%%%%%%%%%%%%%%%%%%%%%%%%%%%%%%%%%%%%%
%%%%%%%%%%%%%%%%%%%%%%%%%%%%%%%%%%%%%%%%%%%%%%%
%%%%%%%%%%%%%%%%%%%%%%%%%%%%%%%%%%%%%%%%%%%%%%%

\clearpage
\section{Conclusion}
\label{sec_conclusion}

In this work we have constructed all flux vacua 
for a type IIB orientifold compactification on the mirror octic
with flux numbers $N_{\rm flux}\leq 10$. 
To achieve this, we  developed and applied techniques for determining
a finite set of flux configurations for each flux number.
Our results agree in most parts with the literature.

%%%%%%%%%%%%%%%%%%%%%%%%%%%%%%%%%%%%%%%%%%%%%%%
%%%%%%%%%%%%%%%%%%%%%%%%%%%%%%%%%%%%%%%%%%%%%%%

\subsubsection*{Results}

Let us  summarize our main results:
\begin{itemize}

\item The orientifold projection for our model is such that the 
D3-brane tadpole contribution of orientifold planes and D7-branes is $Q_{\rm D3}=8$.
(We refer to \cite{Moritz:2023jdb} for the techniques to determine the projection.)
We constructed all flux configurations with $N_{\rm flux}\leq10$ and, hence, we have determined all 
flux vacua for our model consistent with the tadpole cancellation condition.

\item The distribution of vacua on the complex-structure moduli space has been 
illustrated in figures~\ref{fig_minima} and \ref{fig_minima_all}. 
We observed that the density of vacua around the conifold point 
is higher than in other regions, in agreement with  \cite{Denef:2004ze}.

\item In section~\ref{sec_bounds} we have derived bounds on the 
NS-NS and R-R three-form fluxes in finite regions of the complex-structure moduli 
space. Here finite means that the eigenvalues of the Hodge-star operator are finite, 
which excludes boundary points such as the large-complex-structure  or conifold point.
These bounds lead to a finite set of flux choices for these regions, and we emphasize that
the bounds are applicable 
for  arbitrary $h^{2,1}_-$.

\item For the boundary regions  near the large-complex-structure and conifold point 
we explained in sections~\ref{sec_region_lcs} and \ref{sec_region_coni}
how all relevant flux choices can be determined. This approach 
can also be applied to other examples, for instance the large-complex-structure limit 
of the isotropic torus. 

\item Our results for the cumulative number of flux vacua 
$  \mathcal N (N_{\rm flux}\leq N_{\rm max})$
are shown in table~\ref{tab_number_vacua_2} and have been plotted in 
figure~\ref{fig_number_vacua}. Our results are in agreement with the 
estimate based on the counting formula of  \cite{Ashok:2003gk,Denef:2004ze}.

\item For the Landau-Ginzburg point of the mirror octic the Hodge-star operator simplifies,
as shown in section~\ref{sec_region_LG}. 
We were able to determine all flux vacua for flux numbers $N_{\rm flux}\leq 10$
for the LG point, and the  scaling of the number of vacua with $N_{\rm max}$ matches the results obtained in \cite{DeWolfe:2004ns}. 
However, different from \cite{DeWolfe:2004ns}, we also find vacua with $W_0=0$ at the 
LG point. 

\item The distribution of the value of the superpotential 
at the minimum  $e^{\mathcal K_{\rm cs}}\lvert W_0\rvert^2$ 
has been shown in figure~\ref{fig_w0_1}. We did not find an analytic expression for 
this distribution, but it is not a normal distribution. 
On the other hand, as shown in figure~\ref{fig_w0_2},
for $e^{\mathcal K_{\tau}+\mathcal K_{\rm cs}}\lvert W_0\rvert^2$
we do find a normal distribution 
for which the mean and standard deviation scale linearly with 
the flux number. Our findings are different from  \cite{Ebelt:2023clh}, 
as explained on page~\pageref{page_ebelt}.

\item In table~\ref{table_fraction} we collected information on what fraction of flux vacua
is in principle accessible when sampling flux choices from a finite box. Our results 
for $h^{2,1}_-=1$ 
show that for obtaining the majority of flux vacua via a sampling approach,
 the range from which flux quanta are sampled has to be 
 large.

\end{itemize}

%%%%%%%%%%%%%%%%%%%%%%%%%%%%%%%%%%%%%%%%%%%%%%%
%%%%%%%%%%%%%%%%%%%%%%%%%%%%%%%%%%%%%%%%%%%%%%%

\subsubsection*{Outlook}

We also comment on future directions and applications of our results:
\begin{itemize}

\item The techniques developed in this work can  be applied to settings with more moduli. 
A natural next step is to consider an example with two complex-structure moduli or 
to lift the mirror-octic orientifold to an F-theory setting. 

\item The desert point has been proposed as a candidate for 
the inner-most point of  moduli space.
In \cite{vandeHeisteeg:2022btw} the desert point for the mirror quintic has been identified 
with the Landau-Ginzburg point and, due to the similarity with the quintic, we 
expect that this is true also for the mirror octic.
It is then curious to note that all vacua with $W_0=0$ are located at the LG point, and it 
would be interesting to compute the  distances to the other vacua and compare
them to the arguments made in \cite{Plauschinn:2021hkp}.

\item The bounds derived in section~\ref{sec_bounds} are helpful 
for constraining the sampling region for statistical 
searches of vacua, especially for models with a large number of complex-structure moduli.  
It would be interesting to implement these bounds for a concrete model.

\end{itemize}

%%%%%%%%%%%%%%%%%%%%%%%%%%%%%%%%%%%%%%%%%%%%%%%
%%%%%%%%%%%%%%%%%%%%%%%%%%%%%%%%%%%%%%%%%%%%%%%
%%%%%%%%%%%%%%%%%%%%%%%%%%%%%%%%%%%%%%%%%%%%%%%
%%%%%%%%%%%%%%%%%%%%%%%%%%%%%%%%%%%%%%%%%%%%%%%
%%%%%%%%%%%%%%%%%%%%%%%%%%%%%%%%%%%%%%%%%%%%%%%
%%%%%%%%%%%%%%%%%%%%%%%%%%%%%%%%%%%%%%%%%%%%%%%
%%%%%%%%%%%%%%%%%%%%%%%%%%%%%%%%%%%%%%%%%%%%%%%
%%%%%%%%%%%%%%%%%%%%%%%%%%%%%%%%%%%%%%%%%%%%%%%
%%%%%%%%%%%%%%%%%%%%%%%%%%%%%%%%%%%%%%%%%%%%%%%

\subsection*{Acknowledgments}

We thank
Brice Bastian,
Thomas Grimm,
Damian van de Heisteeg,
Sven Krippendorf,
Jeroen Monnee,
Jakob Moritz,
Andreas Schachner, and
Mick van Vliet
for very helpful discussions. 
We furthermore thank 
Elisa Chisari
for help with setting-up the cluster computations at Utrecht University.
The work of EP is supported by a Heisenberg grant of the
\textit{Deutsche Forschungsgemeinschaft} (DFG, German Research Foundation) 
with project-num\-ber 430285316.
The work of LS is supported by the Dutch Research Council (NWO) through a Vici grant.

%%%%%%%%%%%%%%%%%%%%%%%%%%%%%%%%%%%%%%%%%%%%%%%
%%%%%%%%%%%%%%%%%%%%%%%%%%%%%%%%%%%%%%%%%%%%%%%
%%%%%%%%%%%%%%%%%%%%%%%%%%%%%%%%%%%%%%%%%%%%%%%
%%%%%%%%%%%%%%%%%%%%%%%%%%%%%%%%%%%%%%%%%%%%%%%
%%%%%%%%%%%%%%%%%%%%%%%%%%%%%%%%%%%%%%%%%%%%%%%
%%%%%%%%%%%%%%%%%%%%%%%%%%%%%%%%%%%%%%%%%%%%%%%
%%%%%%%%%%%%%%%%%%%%%%%%%%%%%%%%%%%%%%%%%%%%%%%
%%%%%%%%%%%%%%%%%%%%%%%%%%%%%%%%%%%%%%%%%%%%%%%
%%%%%%%%%%%%%%%%%%%%%%%%%%%%%%%%%%%%%%%%%%%%%%%
%%%%%%%%%%%%%%%%%%%%%%%%%%%%%%%%%%%%%%%%%%%%%%%
%%%%%%%%%%%%%%%%%%%%%%%%%%%%%%%%%%%%%%%%%%%%%%%
%%%%%%%%%%%%%%%%%%%%%%%%%%%%%%%%%%%%%%%%%%%%%%%
%%%%%%%%%%%%%%%%%%%%%%%%%%%%%%%%%%%%%%%%%%%%%%%
%%%%%%%%%%%%%%%%%%%%%%%%%%%%%%%%%%%%%%%%%%%%%%%
%%%%%%%%%%%%%%%%%%%%%%%%%%%%%%%%%%%%%%%%%%%%%%%
%%%%%%%%%%%%%%%%%%%%%%%%%%%%%%%%%%%%%%%%%%%%%%%
%%%%%%%%%%%%%%%%%%%%%%%%%%%%%%%%%%%%%%%%%%%%%%%
%%%%%%%%%%%%%%%%%%%%%%%%%%%%%%%%%%%%%%%%%%%%%%%

\clearpage
\nocite{*}
\bibliography{references}
\bibliographystyle{utphys}

%%%%%%%%%%%%%%%%%%%%%%%%%%%%%%%%%%%%%%%%%%%%%%%
%%%%%%%%%%%%%%%%%%%%%%%%%%%%%%%%%%%%%%%%%%%%%%%
%%%%%%%%%%%%%%%%%%%%%%%%%%%%%%%%%%%%%%%%%%%%%%%
%%%%%%%%%%%%%%%%%%%%%%%%%%%%%%%%%%%%%%%%%%%%%%%
%%%%%%%%%%%%%%%%%%%%%%%%%%%%%%%%%%%%%%%%%%%%%%%

\end{document}